\documentclass[]{aa}

\usepackage{graphicx}
\usepackage{txfonts}
\usepackage{lipsum}
\usepackage{subcaption}         
                                
\usepackage{lscape}             
                                
\usepackage{placeins}

\usepackage[colorlinks, linkcolor=black, anchorcolor=green, citecolor=blue]{hyperref}
\usepackage{grffile}
\usepackage{tikz}

\usepackage{ulem}
\usepackage{url}
\usepackage{threeparttable}
\usepackage{xcolor}
\usepackage{soul}
\usepackage{booktabs}
\usepackage{natbib}
\usepackage{amsmath}
\usepackage{mathrsfs}

\usepackage{rotating}

\renewcommand{\thesubfigure}{\Roman{subfigure}}
\usepackage{multirow}
\usepackage{graphbox}

\begin{document}

\title{
Spatially resolved thermal dust emission in the L1157 outflow reveals grain-driven molecular enrichment
}

 \titlerunning{Dust Resolved in L1157 Shocks}

   \author{Siyi Feng\inst{1}\fnmsep\thanks{Corresponding author: syfeng@xmu.edu.cn}
        \and Hauyu Baobab Liu\inst{2,3}
        \and Yang Lu\inst{4}
        \and Qiancheng Yang\inst{1}
        \and Sheng-Yuan  Liu\inst{5}
        \and Paola Caselli\inst{6}
        \and Zhi-Yu~Zhang\inst{7,8}
        \and Shuting Lin\inst{1}
        \and Xuejian Jiang\inst{4}
        \and Sihan Jiao\inst{9,10}
        \and Linjing Feng\inst{9,11}
        \and Donghui Quan\inst{12}
        \and Fujun Du\inst{13,14} 
        \and Yuanzhen Xiong\inst{11,9}
        }

\authorrunning{Feng et al.}

   \institute{Department of Astronomy, Xiamen University, Zengcuo'an West Road, Xiamen, 361005
   \and Department of Physics, National Sun Yat-Sen University, No. 70, Lien-Hai Road, Kaohsiung City, 80424
   \and Center of Astronomy and Gravitation, National Taiwan Normal University, Taipei, 116
   \and Research Center for Computational Earth and Space Science, Zhejiang Laboratory, Hangzhou, 311100
   \and Institute of Astronomy and Astrophysics, Academia Sinica, 11F of Astronomy-Mathematics Building, AS/NTU No. 1, Section 4, Roosevelt Road, Taipei, 106319
   \and Max-Planck-Institut f\"ur Extraterrestrische Physik, Gie{\ss}enbachstra{\ss}e 1,  D-85748,  Garching bei M\"unchen, Germany
   \and School of Astronomy and Space Science, Nanjing University, Nanjing, 210093
   \and Key Laboratory of Modern Astronomy and Astrophysics (Nanjing University), Ministry of Education, Nanjing, 210093
   \and National Astronomical Observatories, Chinese Academy of Sciences, 20A Datun Road, Chaoyang District, Beijing, 100012
   \and Max Planck Institute for Astronomy, K\"onigstuhl 17,  Heidelberg, D-69117, Germany 
   \and University of Chinese Academy of Sciences, Beijing, 100049
   \and Department of Physics, School of Mathematics and Physics, Xi’an Jiaotong-Liverpool University, 111 Ren’ai Road, Suzhou Dushu Lake Science and Education Innovation District, Suzhou Industrial Park, Suzhou, 215123
   \and Purple Mountain Observatory and Key Laboratory of Radio Astronomy, Chinese Academy of Sciences, Nanjing, 210033
   \and School of Astronomy and Space Science, University of Science and Technology of China, Hefei, 230026}

   \date{}

  \abstract 
  {Protostellar outflow shocks reshape local dust properties and molecular chemistry. The L1157 outflow is an archetypal chemically rich shocked region, but the thermal dust associated with its successive shocks has remained unresolved because  molecular-line contamination obscures the broadband continuum.}
   {We aim to resolve the  dust emission toward L1157 B0-B1-B2 {and establish observational constraints on the relationship between the dust evolution and shock-driven chemistry. }
   }
  {We {obtained} new James Clerk Maxwell Telescope (825\text{--}906\,$\mu$m) spectral-line observations and Submillimeter Array (1.1\text{--}1.4\,mm)  continuum observations toward  L1157 B0-B1-B2,  probing {spatial scales} from 0.4\,pc to 1200\,au. After {removing molecular-line contamination on a pixel-by-pixel basis, we derived the dust temperature and density profile and managed to constrain the dust spectral index using continuum data from 70\,$\mu$m to 1.3\,mm. Combining  with previous $\rm NH_3$ observations,  these data were interpreted using a newly developed physicochemical shock model.}  
  }
   {{The line-corrected continuum maps reveal the dust distribution across successive shocks in the southern lobe of L1157.}
   At $\rm 10^4$\,au resolution, we recovered the dust temperature from the outflow cavities to the  protobinary  envelope. The dust opacity index  ($\beta\approx$1.8\text{--}2.3)  indicates that grains have not grown to millimeter sizes throughout the shocked regions. 
   At $\rm 10^3$\,au  resolution, the dust emission resolves into compact clumps along the precessing jet, whereas gaseous $\rm NH_3$ {peaks at} the shocked fronts, reaching abundances of $\sim10^{-5}$ with respect to $\rm H_2$, even where the 0.85 and 1.3\,mm  dust emission is detected at only 3\text{--}5$\sigma$.
  Our modeling shows that $\rm NH_3$ forms predominantly on grain surfaces and is released through shock-induced sputtering, with the highest abundances occurring where 
  {post-shock re-adsorption remains inefficient.} 
  }
{
Spatially resolved dust continuum imaging provides a direct observational probe of grain evolution, highlighting the fundamental role of dust evolution in shaping the chemistry of protostellar shocks.}

   \keywords{astrochemistry --
                 ISM: jets and outflows --
                 (ISM:) dust, extinction--
                 stars: protostars--
                 (stars:) circumstellar matter
               }

   \maketitle

\section{Introduction}

Protostellar shocks are crucial for understanding how young stellar objects (YSOs) inject energy and momentum into their natal molecular clouds. Induced by collimated jets that are driven by accretion, the protostellar outflow shocks compress, heat, and accelerate the ambient gas, shaping both the kinematics and the chemistry of the interstellar medium \citep[ISM, e.g.,][]{bally16}. They provide elevated temperatures that enable otherwise inaccessible gas-phase reactions and induce grain processing, such as grain-grain collision and sputtering \citep{caselli97}, which releases refractory and icy mantle species into the gas phase \citep[e.g.,][]{bachiller97,jimenez05,arce07,viti11,williams13}.

Dust grains, while comprising only 1\% of the interstellar mass, play a vital role in the physical and chemical evolution throughout star and planet formation.
In dense molecular cloud cores, they act as chemical reservoirs where atoms and molecules adsorb, while their catalytic surfaces efficiently dissipate excess reaction energy and enhance exothermic processes such as the hydrogenation of C, N, O, and CO. This results in the production of such species as  $\rm H_2$ and $\rm NH_3$, along with more complex molecules as $\rm H_2CO$ and $\rm CH_3OH$ \citep[e.g., ][]{cazaux04,vD14,jorgensen20}. Moreover, they regulate the temperature structure through extinction and thermal emission, thereby extending molecular lifetimes by shielding against UV radiation, enabling freeze-out and reformation, moderating ion chemistry, and stabilizing the thermal environment. These processes support  further gas-phase formation of long carbon chains ($\rm HC_{2n+1}N$, $\rm C_{2n+1}S$, $\rm C_{2n}H$, n=1,2,3...)  {as well as structurally more complex molecules}
\citep[e.g., ][]{tielens05}.
 In shock regions, dust grains are subject to thermal processing, fragmentation, and destruction. These changes can lead to significant variations in the grain size distribution and composition, which, in turn, affect the dust opacity, emissivity index, ion-neutron fraction, and dust-to-gas ratio \citep[e.g., ][]{guillet11,zhao18,tsukamoto23a,lebreuilly23}. Understanding the evolution of dust in shocked regions is therefore essential for interpreting observed molecular abundances and for modeling the radiative transfer of both continuum and line emission.

Despite their importance, direct detections of thermal dust continuum emission impacted by the protostellar shocks remain rare.
 Only a few recent studies have reported detections of compact dust emission associated with {the outflow cavity wall}, such as {the dust continuum in the mm wavelength detected} toward {Corona Australis (CrA)} IRS7B \citep{sabatini24} and L1551 IRS5 \citep{sabatini25}, {as well as warm dust detected at infrared band toward BHR71-IRS1 \citep{tychoniec26}}; however, their fields of view have only enabled the dust grain variation to be traced within 2000\,au from the protostars. Consequently, 
 the properties of shocked dust, such as its temperature and density structure, mass, and potential changes in emissivity due to shocks, remain poorly constrained.

The Class 0 {protobinary} L1157-mm ({\citealt{tobin22}; } $d\sim$352\,pc, \citealp{zucker19}; $\sim340$ pc, \citealt{sharma20}) drives a chemically rich bipolar outflow. Its southern lobe features several successive bright shocks, most notably  {B0, B1} and B2 \citep[e.g., ][]{bachiller97,bachiller01}, which are separated by $\sim$50\arcsec ~(0.1\,pc) and differ by $\sim$1000\,yr in dynamical age \citep[e.g., ][]{gueth98, podio16, sharma20}. These shocked regions exhibit {rich molecular inventories, including both simple species and complex organic molecules (COMs; defined as containing $\ge$6 atoms, e.g., \citealt{herbst09}).
These detections, particularly of species with key prebiotic relevance, indicate that active grain processing and shock-induced chemistry are taking place \citep[e.g.,][]{codella10,codella12,lefloch10,lefloch12,busquet13,podio14,mendoza14,holdship16,lefloch17,codella20}.}

{
Although the pioneering studies of \citet{shirley00}, \citet{chini01}, and \citet{gueth03} established the presence of submillimeter (submm) and millimeter (mm) emission associated with the L1157 outflow, the available observational capabilities did not permit a direct, spatially resolved isolation and characterization of the thermal dust component, owing to the difficulty of quantitatively separating molecular-line contamination from the bolometric emission.
Interferometric observations can, in principle, directly isolate the thermal dust component by simultaneously separating line and line-free channels and measuring the continuum spectral index. However, previous interferometric studies at 2--3\,mm and centimeter wavelengths have not detected compact continuum emission associated with the shocked cavities on scales below $\sim4000$\,au \citep[e.g.,][]{burkhardt16,feng20a,feng22}, leaving the dust properties in the shocked gas of this region largely unconstrained.}

We present new observations of L1157 outflow at submm and mm wavelengths in Section~\ref{sec:obs}. In Section~\ref{sec:obsresult}, we provide the first direct, line-corrected, and spatially resolved characterization of the thermal dust emission toward the L1157 B0–B1–B2 shocks, yielding new constraints on dust survival and evolution in protostellar outflow shocks in Section \ref{sec:profile}. Finally, in Section~\ref{sec:abundnace}, by combining the derived dust properties with a new physicochemical model and our previous $\rm NH_3$ observations, we demonstrate how grain evolution drives molecular enrichment and dust-gas interactions over scales ranging from 0.4\,pc to 1200\,au.

\section{Observations and data reduction} \label{sec:obs}

\subsection{Submillimeter Array}
Using the Submillimeter Array (SMA), we observed the B1-B2 region at 1.1\text{--}1.4\,mm in the compact configuration from June to July in 2017. A two-point mosaic was centered at $\rm 20^h39^m09^s.500$, $\rm 68^{\circ}01^{'}10^{''}.50$ and $\rm 20^h39^m11^s.358$, $\rm 68^{\circ}00^{'}51^{''}.00$ (J2000). 
The array consisted of seven or eight antennas with baselines ranging from 16 to 76\,m at different dates. Observations on B1 and B2 shared the same bandpass calibrators (3C273 or 3C454.3) and flux calibrators (Callisto and MWC349a).
The zenith opacities,  measured with water vapor monitors mounted on the James Clerk Maxwell Telescope (JCMT), were satisfactory throughout all tracks with $\tau_{225\,\rm GHz}\sim0.05\text{--}0.3$.
The system temperature was 100\text{--}200\,K at 230\,GHz and 200\text{--}400\,K at 255\,GHz.
 Further details of these observations are described in {\citet{feng20a}.}

From October to November {in}  2024, we observed the B0 region, also in compact configuration at 1.1\text{--}1.4 \,mm, using another two-point mosaic centered at $\rm 20^h39^m07^s.413$, $\rm 68^{\circ}01^{'}53^{''}.98$ and $\rm 20^h39^m08^s.518$, $\rm 68^{\circ}01^{'}32^{''}.23$ (J2000).
The baselines range from 16\,m to 77\,m with five or six {antennas} at different dates. 
The bandpass calibrators were BLLAC, 3C84, or MWC349, {and} the flux calibrators were Neptune or Titan. The zenith opacities during all tracks were stable ($\tau_{225\,\rm GHz}\sim0.2\text{--}0.3$), and the system temperatures ranged from 100\text{--}400\,K at 230\,GHz to 200\text{--}500\,K at 255\,GHz.

Using the  SWARM \citep[SMA Wideband Astronomical ROACH2 Machine, ][]{primiani16} correlator, we sampled our observations with a spectral resolution of 0.140\,MHz. We employed the {dual-receiver (Rx230 and Rx240)}, single-polarization, double-sideband mode.
 {Following the SMA/SWARM correlator terminology, we refer to the 2\,GHz digital sub-bands processed by individual ROACH2 boards as spectral ``chunks.'' 
For both the 2017 and 2024 observations, the receivers were tuned so that 230.0\,GHz and 255.8\,GHz were placed at the center of spectral chunk {\it s1} in the upper sideband of Rx230 and Rx240, respectively. 
The 2017 observations were conducted during the partial deployment phase of SWARM, resulting in an effective bandwidth of 24.0\,GHz and frequency coverage of 214.0\text{--}220.0, 230.0\text{--}236.0, 239.8\text{--}245.8, 255.8\text{--}261.8\,GHz. 
Following the full SWARM upgrade, the 2024 observations achieved a larger effective bandwidth of 35.8\,GHz with the same receiver tuning, extending the frequency coverage to 210.0\text{--}220.0, 230.0\text{--}245.8, 255.8\text{--}265.8\,GHz.}
 The full width at half maximum (FWHM) of the primary beam is $\sim$52\arcsec~ at 237.9\,GHz. 
 Further technical descriptions of the SMA and its calibration schemes can be found in \citet{ho04}.

The data reduction was performed using the MIR software package \citep{scoville93, qi03b}\footnote{The MIR package was originally developed for the Owens Valley Radio Observatory, and is now adapted for the SMA, \url{http://cfa-www.harvard.edu/~cqi/mircook.html}}, followed by imaging with MIRIAD using natural weighting \citep{sault95}.  Line-free channels were averaged to produce continuum maps. The resulting 1.1\text{--}1.4\,mm continuum achieves a sensitivity of $\rm \sigma = 0.4\,mJy\,beam^{-1}$ with a {synthesized beam of  $\rm 3.7\arcsec\times 3.1\arcsec$ (P.A.$\sim3^\circ$).}

\subsection{James Clerk Maxwell Telescope (JCMT)}

Using JCMT with the 16-receptor array receiver Heterodyne Array Receiver Program \citep[HARP,][]{buckle09}, we carried out a total of 31\,hrs line-imaging survey toward L1157  from April to December {in} 2023 (ID: M22BF001). 
 The Auto-Correlation Spectral Imaging System (ACSIS) served as the backend spectrometer. A total of 18 frequency setups were used to cover the frequency range of  329.9\text{--}362.4\,GHz. Each setup employed two 1000\,MHz-wide spectral windows to cover different frequency ranges, with a channel resolution of  0.488\,MHz. 

At 350\,GHz, the HARP beam has a FWHM of $\sim$14.5\arcsec. Observations were carried out in raster scan mode with a grid spacing of 1/4 array ($7.2761\arcsec$), resulting in a $3\arcmin\times3\arcmin$ map centered on $\rm 20^h39^m09^s.912$, $+68^{\circ}01^{'}30^{''}.28$ (J2000), which covers the protobinary disk as well as the B0, B1, and B2 shocked regions. We employed position-switching mode, with an off-position at $\rm 20^h44^m01^s.900$, $+68^{\circ}31^{'}15^{''}.00$ (J2000).

Pointing and focus calibrations were performed using standard sources, including CRL618, 21282+5050, CRL2688, NGC7027, 21318+5631, TCep, HD235858. The flux calibration was performed using standard line calibrators, including CRL618, W75N, 16293-2422, and G34.3. We estimated the uncertainty on the absolute flux calibration to be $\sim$10\%. All observations were carried out under good weather conditions, with $\tau_{225\,\rm GHz}$ in the range 0.05\text{--}0.10.

The data reduction was performed using the Starlink software suite\footnote{\url{https://starlink.eao.hawaii.edu/starlink}} and the ORAC-DR pipeline \citep{currie14,jenness15}, adopting the “REDUCE SCIENCE NARROWLINE” recipe with the default parameters. A first-order baseline was fitted and subtracted from each spectrum using line-free channels.

{The final line data cubes achieve a $1\sigma$ rms noise level in antenna temperature ($T_A^*$) ranging from 0.01 to 0.04\,K across the different frequency setups. A preliminary identification of the most prominent molecular emission is presented in Appendix~\ref{app:linejcmt}, including {the J=3\text{--}2 transitions of the CO isotopologues}, SiO (8\text{--}7), CS (7\text{--}6), multiple transitions of SO, $\rm CH_3OH$, and $\rm H_2CO$, as well as hyperfine structures of CN (N=3-2) and HCN\,(J=4\text{--}3).  
{Numerous additional transitions and several unidentified features were also detected. A comprehensive line census, together with detailed chemical and kinematic analyses, is beyond the scope of this paper and will be presented in a dedicated follow-up study.} Since the detected molecular emission is generally extended relative to the JCMT beam, the antenna temperature ($T_A^*$) was first converted to the source-coupled radiation temperature using the forward spillover and scattering efficiency $\eta_{\rm fss}=0.71$ \citep{drabek12}, and subsequently converted into flux density units ($\rm Jy\,beam^{-1}$).
}

\subsection{Archival data}

{
To obtain bolometric measurements at submm wavelengths, we also incorporated archival \textit{Planck} 850\,$\mu$m data \citep{planck15}\footnote{\textit{Planck} is an ESA science mission with instruments and contributions directly funded by ESA Member States, NASA, and Canada, http://www.esa.int/Planck} and JCMT/SCUBA-2 850\,$\mu$m data (Program IDs: MJLSG40, M17AP073), together with \textit{Herschel} Gould Belt Survey (HGBS) data at 70, 160, 250, 350, and 500\,$\mu$m \citep{andre10,difrancesco20}.

To remove molecular line contamination from the broadband bolometric measurements, we made use of the available archival spectral-line data cubes. These included observations toward the central protobinary and envelope system from the COPS-DIGIT-FOOSH (CDF) \textit{Herschel} spectroscopy archive at 70 and 160\,$\mu$m \citep{green16}, as well as data from the CO in Protostars (COPS) program at 250, 350, and 500\,$\mu$m \citep{yang18}. Moreover, we used observations targeting the B1 cavity and shocked region from the Chemical HErschel Surveys of Star-forming Regions (CHESS) program at 70 and 160\,$\mu$m \citep{codella10,lefloch10}.

  }

\section{Results}\label{sec:obsresult}
To obtain the dust continuum distribution while minimizing line contamination in the bolometric measurements, we adopted the following wavelength-dependent approach:

At 850\,$\mu$m, we recovered the spatially filtered flux in the SCUBA-2 data by combining it with {matching-frequency} \textit{Planck} observations using the J-comb algorithm\footnote{Because SCUBA-2 bolometric observations lack line-free channels for separating astronomical emission from atmospheric fluctuations, the map-making process suppresses large-scale correlated signals to remove sky emission, resulting in attenuation of diffuse structures on large angular scales. In contrast, HARP spectral-line observations preserve extended emission through baseline subtraction and position-switching techniques. The J-comb method restores the missing low spatial-frequency emission while retaining the high-resolution information from JCMT.}. {The J-comb algorithm is described in detail by} \citet{lin16,lin17} and \citet{jiao22}. 

The resulting combined 850\,$\mu$m map achieves an angular resolution of $\sim14.5\arcsec$.
 Since the HARP observations cover the same frequency range and have a comparable angular resolution to the SCUBA-2 and \textit{Planck} combined data, we followed the approach of \citet{smith19} to subtract the summed spectral line intensities from the bolometric  map on a pixel-by-pixel basis.

At 1.3\,mm, we constructed a high spatial resolution dust continuum image, five times finer than that of JCMT,  using line-free spectral channels from our broadband, high-sensitivity SMA line imaging survey carrying between 2017 and 2024.

\begin{figure}
  \centering
        \begin{tabular}{lc}
        \rotatebox{90}{\parbox{8cm}{\centering $\Delta$ Dec.($\arcsec$) }}
&\includegraphics[clip, trim=0.0cm 0.0cm 0.00cm 0.0cm,height=8cm]{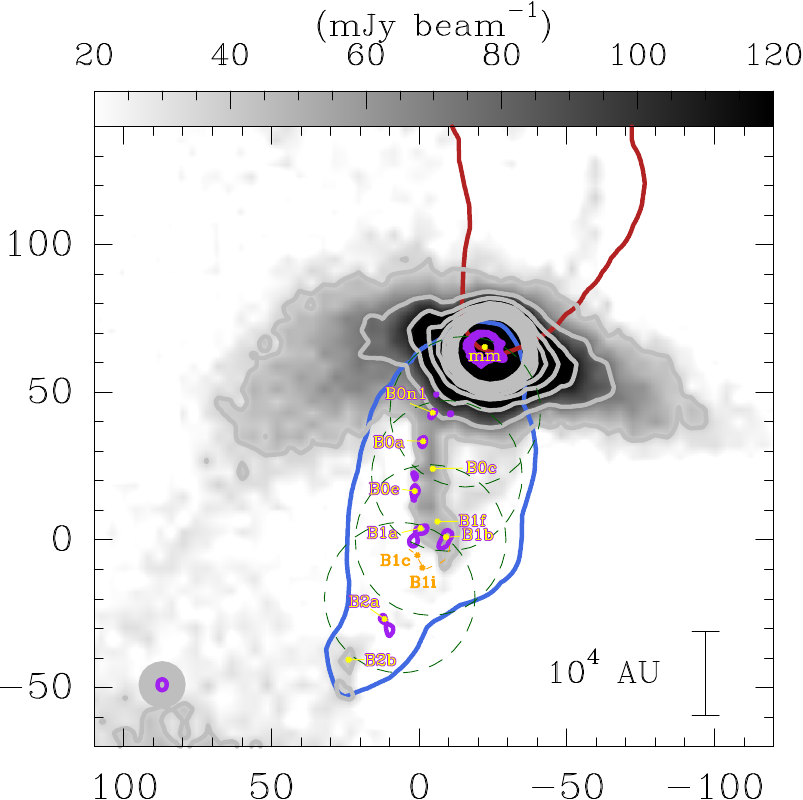}\\
&$\Delta$R.A.($\arcsec$) 
\end{tabular}
\caption{
Line-contamination-corrected  continuum emission at 0.85\,mm  shown in grayscale, obtained by combining the JCMT SCUBA-2 and \textit{Planck} data {after removing the contribution from molecular-line emission observed with JCMT-HARP. 
To facilitate comparison with previous studies, the (0,0) offset position is set to the phase center adopted in the NOEMA and VLA single-pointing observations of  \citet{feng20a,feng22}: $\rm RA = 20^h39^m10^s.200$, $\rm Dec = +68^{\circ}01\arcmin10.50\arcsec$ (J2000).}
Gray contours indicate  continuum emission levels from {$\rm 10\sigma$ to $\rm 100\sigma$ in steps of $\rm 10\sigma$ ($\rm \sigma =3.7\,mJy\,beam^{-1}$).} 
Red and blue contours denote  the {integrated CO\,(3\text{--}2) emission above the  $\rm 5\sigma$  level}  over the velocity range of 7.0\text{--}34.0 and -20.0\text{--}-0.6\,$\rm km\,s^{-1}$, {respectively, relative to $\rm V_{lsr}\sim2.6\,km\,s^{-1}$, tracing the red- and blue-shifted outflow lobes. } 
Purple contours represent the SMA 1.3\,mm dust continuum derived from line-free  channels, starting at $\rm 5\sigma$ and increasing in $\rm 5\sigma$ steps ($\rm \sigma = 0.4\,mJy\,beam^{-1}$). 
{The JCMT beam (FWHM = 14.5\arcsec) is shown as a gray filled circle in the lower left corner, while the synthesized SMA beam ($3.7\arcsec \times 3.1\arcsec$, P.A. $\sim3^\circ$) is shown as a purple ellipse.}
Green dashed circles mark the four-point SMA mosaic, with diameters corresponding to the SMA primary beam FWHM at 1.3\,mm. Artifacts near the central protobinary system mm in the northwestern mosaic field caused by noise and/or missing short-spacing  have been masked.
Ten positions with SMA dust continuum detections at the 5\text{--}7$\sigma$ level are {marked in yellow and labeled in  yellow, with purple outline}. 
The B1 shock front is indicated by an orange dashed line, {with orange markers highlighting two representative positions along the front.}
}
\label{fig:source}
\end{figure}

Figure~\ref{fig:source} presents the line-subtracted 850\,$\mu$m JCMT dust continuum emission, which extends from the central protobinary system into both the northern and southern outflow lobes. In the south, the continuum map reveals the successive shocks B0-B1 and B2, with the {eastern} side of B0-B1 and {western} of B2 exhibiting higher signal-to-noise ratios (S/N) than the rest part. In contrast, the northern lobe displays only a narrow, S-shaped filament connecting to the {protobinary} disk–envelope interface. The twisting northern and the southern lobes are consistent with the precessing outflow morphology reported in \citet{podio16}.

{
The zoomed-in 1.3\,mm SMA mosaic {in Figure~\ref{fig:source}} not only reveals strong dust continuum emission toward the central protobinary system (labeled  ``mm''  in the figure), despite its location near the edge of the mosaic, but it also resolves multiple compact dusty clumps distributed along the precessing jet, tracing the eastern cavity walls of B1 and the western cavity wall of B2. Following the naming conventions established in previous molecular-line studies \citep[][]{benedettini07,codella09,burkhardt16,feng20a}, we {labeled} the compact dusty peaks spatially associated as B0a, B0e, B1a, B1b, and B2a (Table~\ref{tab:clump}). The coordinates of these continuum peaks may deviate by up to 5\arcsec~ from the positions reported in the literature according to line emissions. At a  moderate significance level of $\sim5\sigma$, we additionally identified a continuum peak toward B0n1. Although this position lies outside the primary beam coverage of previous single-pointing interferometric molecular-line observations, its presence is independently supported by $>5\sigma$ emission across the line-subtracted 70\text{--}850\,$\mu$m maps. For comparison, we also mark the positions of B0c, B1f, B1c, and B1i identified in previous molecular-line studies.

Comparison between the 1.3\,mm SMA image and the 850\,$\mu$m JCMT-\textit{Planck} map shows broadly consistent dust continuum morphologies. The main exception is the B2 region, where the 1.3\,mm SMA image peaks toward B2a, while the 850\,$\mu$m JCMT-\textit{Planck} map peaks closer to B2b, likely due to missing large-scale flux in the interferometric data. 
More importantly, the 1.3\,mm continuum emission toward the B0-B1 region reproduces the prominent arch-shaped morphology (B1a-B1f-B1b) seen in the 850\,$\mu$m map, while further resolving it into compact dusty clumps. 
}
These dusty clumps spatially coincide with emission peaks of molecules  associated with grain-surface chemistry and shock-induced desorption, including deuterated species such as DCN and HDCO \citep{fontani14}, $\rm CH_2DOH$ \citep{busquet17}, and COMs such as $\rm CH_3CHO$ \citep{codella15}.

In contrast, no continuum emission is detected ($<3\sigma$) toward the B1 shock front in either the {0.4}\,pc-scale 850\,$\mu$m JCMT-\textit{Planck} map {after removing the molecular-line contribution or the 0.1\,pc-scale 1.3\,mm SMA map. This result differs from the previously published bolometric maps \citep{shirley00,chini01,gueth03}, illustrating that molecular-line contamination can substantially alter the apparent morphology of the submm and mm emission (see Appendix~\ref{app:linecontamination} for a detailed comparison).}
{The shock front boundary, indicated by the orange dashed line connecting B1c and B1i, } spatially coincides with the emission peaks of several chemically rich species, including $\rm NH_2CHO$ \citep{codella17}, $\rm CH_3CN$ \citep{codella09}, $\rm CH_3OH$ \citep{benedettini13}, and the phosphorus-bearing species PO and PN \citep{lefloch24}. Since this region is covered by two SMA mosaic pointings, the nondetection is unlikely to {result from} primary-beam attenuation or missing flux, and instead {suggests} an intrinsic lack of detectable  submm and mm dust emission at the shock front.

\begin{table}[tbh]
\small
\caption{Coordinates for the positions labeled in Figure~\ref{fig:source}.
}\label{tab:clump}
\centering
\begin{tabular}{cccc}
\hline
\hline
      &R.A. (J2000) &Decl. (J2000)   &($\Delta$R.A., $\Delta$Decl.)$^a$\\

\hline    
 	    mm       &$\rm 20^h39^m06^s.232$       &$\rm 68^{\circ}02\arcmin15.87\arcsec$     &(-22.3\arcsec, 65.4\arcsec) \\      
      	    B0n1     &$\rm 20^h39^m09^s.373$       &$\rm 68^{\circ}01\arcmin53.60\arcsec$     &(-4.6\arcsec, 43.1\arcsec) \\                      	  		
        	    B0a      &$\rm 20^h39^m09^s.948$       &$\rm 68^{\circ}01\arcmin43.92\arcsec$      &(1.4\arcsec, 33.4\arcsec) \\         
 	    B0c      &$\rm 20^h39^m09^s.363$       &$\rm 68^{\circ}01\arcmin34.62\arcsec$       &(-4.7\arcsec, 24.1\arcsec) \\                 
      	    B0e      &$\rm 20^h39^m10^s.459$       &$\rm 68^{\circ}01\arcmin27.08\arcsec$       &(1.5\arcsec, 16.6\arcsec) \\                   
  	    B1a      &$\rm 20^h39^m10^s.107$       &$\rm 68^{\circ}01\arcmin14.53\arcsec$       &(-0.5\arcsec, 4.0\arcsec) \\                    
      	    B1b      &$\rm 20^h39^m08^s.575$       &$\rm 68^{\circ}01\arcmin11.49\arcsec$       &(-9.1\arcsec, 1.0\arcsec) \\         
	    B1c     &$\rm 20^h39^m10^s.294$       &$\rm 68^{\circ}01\arcmin05.30\arcsec$        &(0.5\arcsec, -5.2\arcsec) \\            
             B1f      &$\rm 20^h39^m09^s.088$       &$\rm 68^{\circ}01\arcmin16.76\arcsec$       &(-6.2\arcsec, 6.3\arcsec) \\    
             B1i     &$\rm 20^h39^m10^s.000$       &$\rm 68^{\circ}01\arcmin01.11\arcsec$        &(-1.1\arcsec, -9.4\arcsec) \\                   
             B2a      &$\rm 20^h39^m12^s.277$       &$\rm 68^{\circ}00\arcmin43.71\arcsec$      &(11.7\arcsec,-26.8\arcsec) \\
             B2b      &$\rm 20^h39^m14^s.452$       &$\rm 68^{\circ}00\arcmin29.99\arcsec$      &(23.9\arcsec,-40.5\arcsec) \\                                                              
\hline
\hline
\multicolumn{4}{l}{Note. a. Relative offsets  are calculated with respect to the reference }\\
\multicolumn{4}{l}{position at  $\rm RA = 20^h39^m10^s.200$, $\rm Dec = +68^{\circ}01\arcmin10.50\arcsec$  (J2000).}\\
\end{tabular}
\end{table}

\section{Discussion}\label{sec:analysis}

\subsection{Temperature and density profile}\label{sec:profile}

\begin{figure*}

  \centering
        \begin{tabular}{lcc}
&\includegraphics[clip, trim=0.0cm 0.0cm 0.0cm 0.0cm,height=5.5cm]{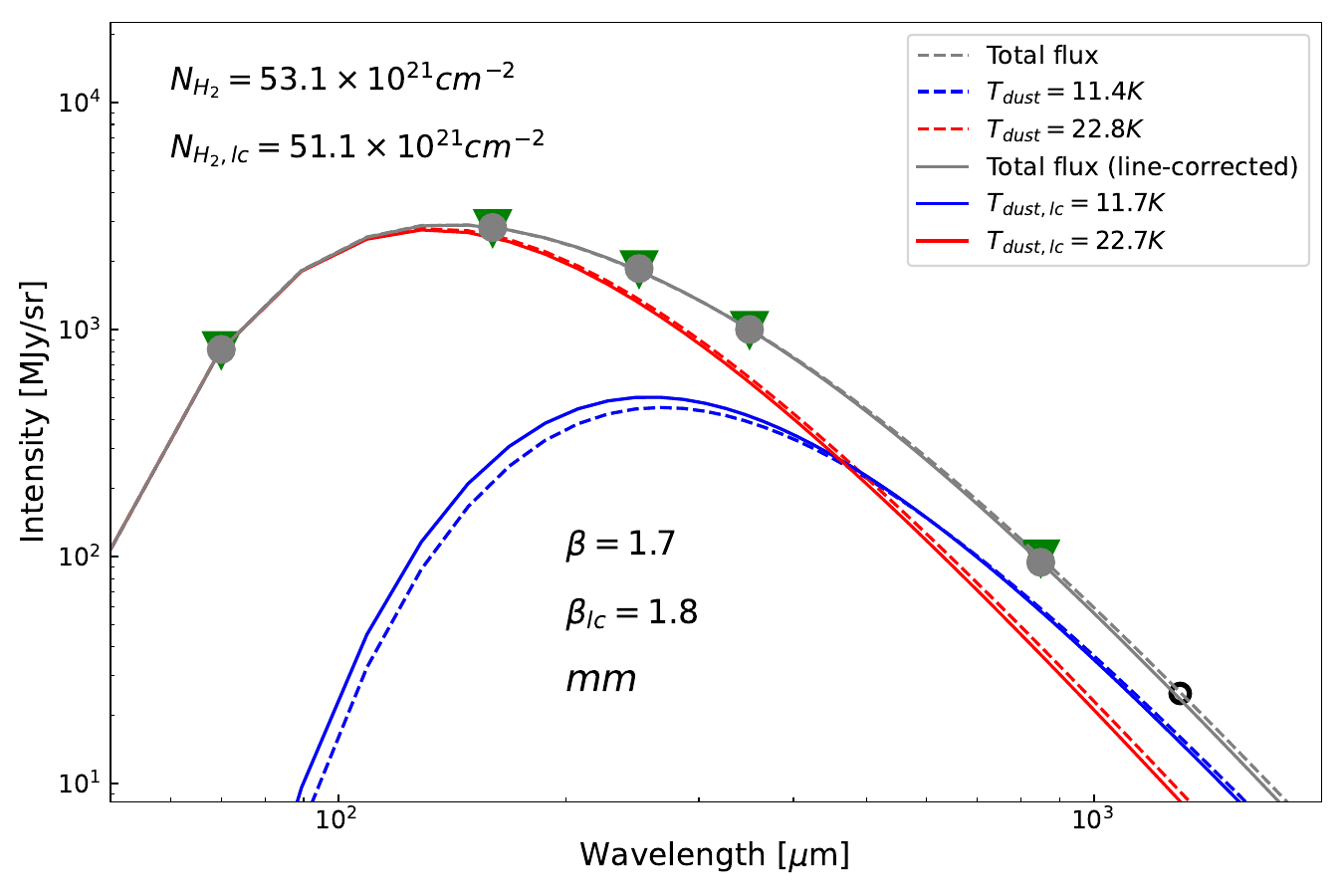}
&\includegraphics[clip, trim=0.0cm 0.0cm 0.0cm 0.0cm,height=5.5cm]{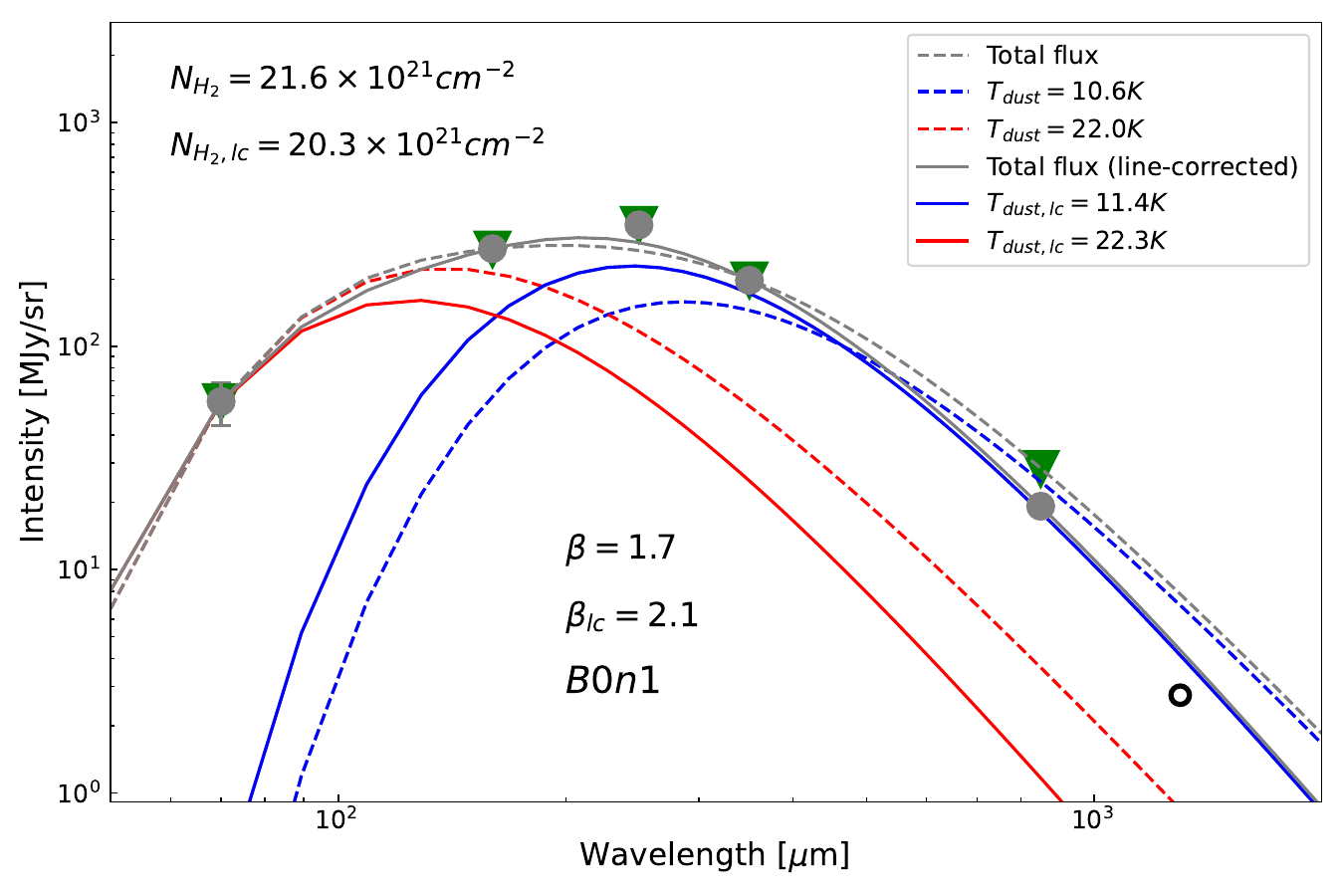}\\
&\includegraphics[clip, trim=0.0cm 0.0cm 0.0cm 0.0cm,height=5.5cm]{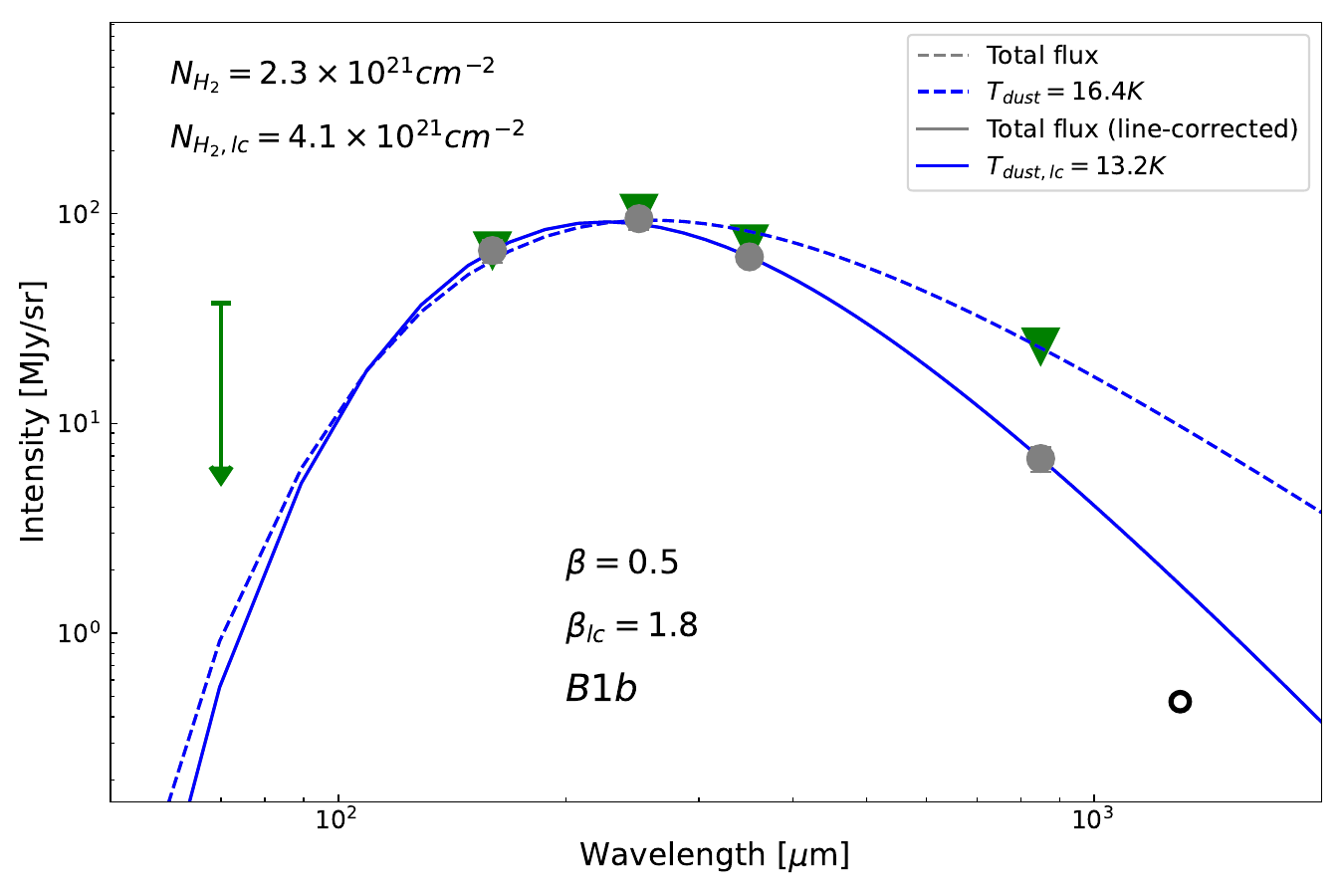}
&\includegraphics[clip, trim=0.0cm 0.0cm 0.0cm 0.0cm,height=5.5cm]{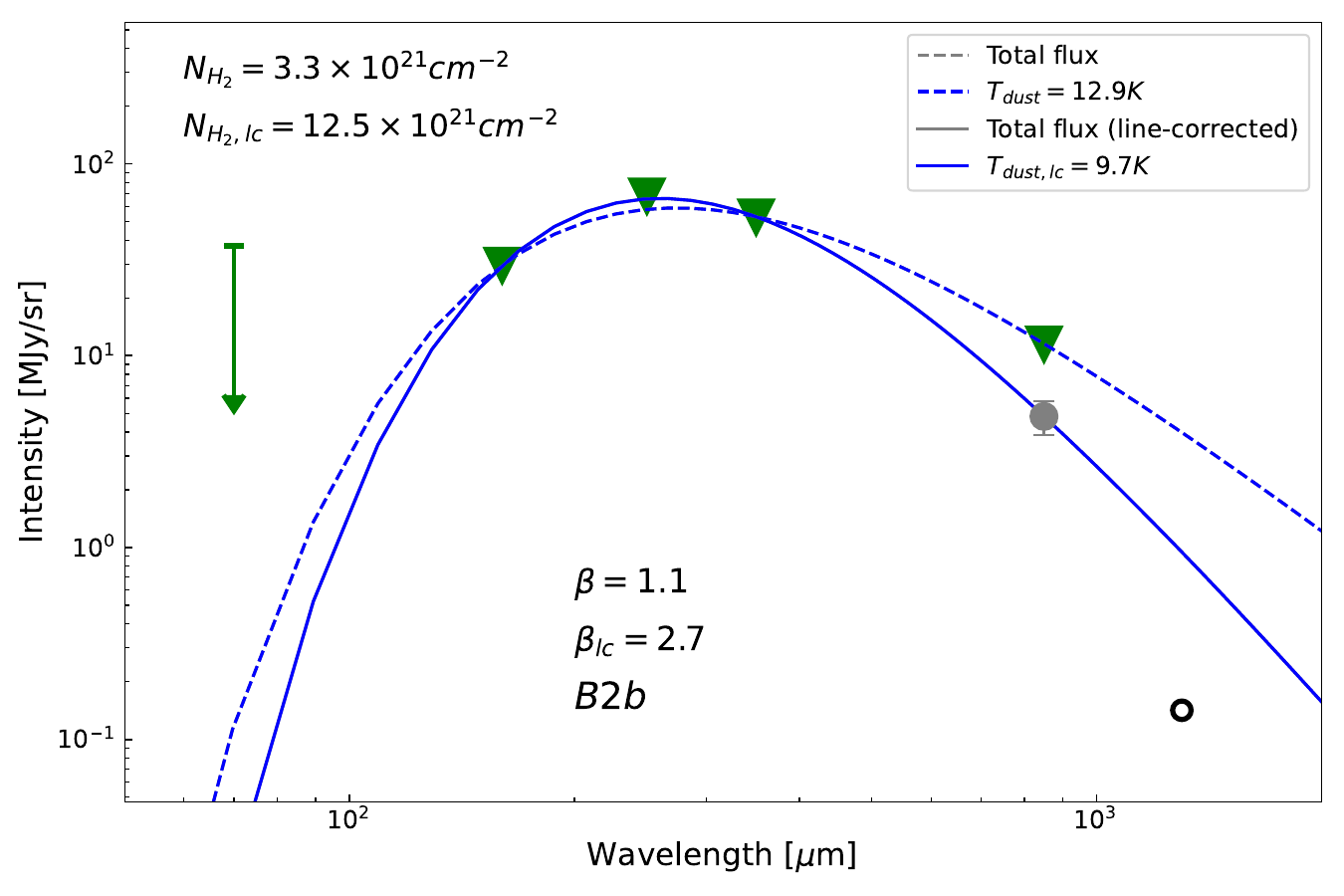}\\
\end{tabular}
\caption{
Two-component SED fits at selected positions. 
The green triangles and gray dots at 70, 160, 250, 350, and 850\,$\mu$m represent the original bolometric fluxes and the line-subtracted continuum fluxes, respectively.
The downward arrow at 70\,$\mu$m indicates a $3\sigma$ upper limit, while the open black circle at 1.3\,mm denotes data affected by missing flux, and both data points are excluded from the fitting.
  All data have been smoothed to a common angular resolution of 24.9\arcsec.
The red and blue curves show the warm and cold components of the two-component SED model, respectively, assuming a gas-to-dust mass ratio of 100,
 while the gray curve represents the total SED obtained by summing both components. 
 Fits using the line-subtracted continuum fluxes are shown as solid curves, whereas fits using the original bolometric fluxes are shown as dashed curves for comparison.
}
\label{fig:sed-point}
\end{figure*}

{
To characterize the dust properties on 0.4 pc scales, we constructed multiwavelength dust continuum maps using \textit{Herschel}, JCMT-\textit{Planck}, and SMA observations. 
Since the available archival spectroscopic observations do not fully match the spatial coverage of the \textit{Herschel} photometric maps at 70, 160, 250, 350, and 500\,$\mu$m, the procedures used to correct for line contamination vary accordingly.

At 70 and 160\,$\mu$m, the PACS photometric observations provide continuous fully sampled imaging at an angular resolutions of 9.4\arcsec and 13.5\arcsec, respectively. 
The corresponding spectroscopic products from the CDF archive \citep{green16} cover a $47\arcsec \times 47\arcsec$ region centered on the protobinary system, 
while the CHESS observations \citep{lefloch10,benedettini12} cover a rotated $42\arcsec \times 42\arcsec$ region centered on B1. 
Although both spectroscopic datasets are fully sampled, the two mapped regions do not overlap in projection.

At 250, 350, and 500\,$\mu$m, the SPIRE photometric maps were also fully sampled, at an angular resolutions of 18.2\arcsec, 24.9\arcsec, and 36.3\arcsec, respectively. 
In contrast, the corresponding SPIRE/FTS spectroscopic observations consist of discrete detector footprints, rather than continuous imaging. 
The COPS spectroscopic products \citep{yang18} cover a circular field centered on the protobinary system with a radius of $\sim108\arcsec$, including B1 and partially covering B2.

To construct spatially continuous line emission maps for the SPIRE bands, positions without direct beam coverage in the SPIRE/FTS observations were first filled using beam-weighted interpolation. 
The \textit{Herschel} photometric and spectroscopic maps were then smoothed to a common angular resolution of 24.9\arcsec, corresponding to the 350\,$\mu$m beam, before subtracting the line contribution from the 70\text{--}350\,$\mu$m bolometric maps. 
The JCMT-\textit{Planck} 850\,$\mu$m and SMA 1.3\,mm continuum images were subsequently smoothed to the same angular resolution. 
At this common resolution, we performed pixel-by-pixel spectral energy distribution (SED)  fitting {over the 0.37\,pc$\rm \times$0.44\,pc mapped region}\footnote{The 100\,$\mu$m data were not obtained simultaneously with the other \textit{Herschel} bands, while the SMA 1.3\,mm continuum data suffer from significant and spatially varying missing flux. Therefore, both datasets were excluded from the SED fitting. We also excluded the 500\,$\mu$m map because its coarser angular resolution (36.3\arcsec) would significantly degrade the spatial information preserved in the 350\,$\mu$m data. Smoothing all images to the 500\,$\mu$m resolution leaves the long-wavelength ($>250\,\mu$m) SED slope nearly unchanged, indicating that both the warm and cold dust-component fits are largely insensitive to the inclusion of this band.}.

At this angular resolution, we can see that the dusty peaks identified in the 1.3\,mm map are not spatially resolved. We therefore extracted the flux densities from representative positions separated by at least one beam width ($\sim25\arcsec$)\footnote{At this resolution, B0a/B0c/B0e are blended with B0n1,  B1a/B1c/B1f/B1i are blended with B1b, and B2a is blended with B2b.}, namely the protobinary system mm, B0n1, B1b, and B2b, and combined the multiwavelength measurements to construct individual SEDs (Figure~\ref{fig:sed-point}).

We define the percentage line contamination $R_{\rm line}$, as the ratio between the integration of the continuum-subtracted line emission, $F_{\rm line}$, to the line-subtracted bolometric flux, $F_{\rm bolo}-F_{\rm line}$,
$R_{\rm line} = 100\times \frac{F_{\rm line}}{F_{\rm bolo}-F_{\rm line}} (\%)$.
 {At 850\,$\mu$m, the JCMT-\textit{Planck} data at an angular resolution of 14.5\arcsec~ reveal substantial molecular-line contamination toward all labeled positions (Table~\ref{tab:contamination}). The bolometric fluxes toward B1c and B1i are dominated by line emission, with line fractions to bolometric flux  of $\sim80\%$ and corresponding $R_{\rm line}$ values of 339\%\text{--}481\%, reducing the significance of the dust continuum detections to below $5\sigma$.
With this pixel-by-pixel correction, we find that the combined contribution from CO and other molecular lines reaches $\sim77\%$ toward B1a, {substantially} extending the  $\sim40\%$ {estimate of} \citet{gueth03}. The higher value {likely reflects our ability to resolve local variations in line contamination that were averaged out in the earlier area-integrated analysis.}

At the angular resolution adopted for the SED fitting (25\arcsec), $R_{\rm line}$
 is only $\sim5\%$ toward the protobinary system, but increases to $\sim48\%$ at B0n1 and reaches $\sim280\%$ and $\sim250\%$ toward the B1 and B2 shocked regions, respectively.
These values correspond to molecular lines contributing approximately 70\% of the bolometric flux toward both B1 and B2, with CO\,(3\text{--}2) accounting for {more than} half of the total line contamination.
This trend is also reflected in the significantly larger number and stronger intensity of spectral lines detected toward B1 and B2 than toward the protobinary position mm, despite the  continuum peaking toward the latter (Figure~\ref{fig:line1}).
In contrast, the line contamination percentage in the \textit{Herschel} 70\text{--}350\,$\mu$m bands remains relatively small, reaching at most {2\%} within the regions covered by the spectroscopic observations (i.e., mm, B0, and B1).
}

Because the HARP line profiles do not show distinct velocity components that would allow for a clear separation of foreground and background material, we first applied a single-temperature graybody model to the multiwavelength line-subtracted bolometric data. However, toward the protobinary system and the B0 region, the observed fluxes at wavelengths longer than 350\,$\mu$m systematically exceed the single-component model predictions (Figure~\ref{fig:sed-point}), suggesting the presence of an additional colder dust component coexisting with warmer protostellar material {at a linear resolution of $\sim\rm 10^4$\,au.}

Motivated by this long-wavelength excess, we performed pixel-by-pixel two-component SED fitting, consisting of a warm protobinary-envelope component and a colder extended environmental component along the line of sight. A gas-to-dust mass ratio of 100 was adopted. 
The model includes five free parameters: the temperatures and column densities of the warm and cold components, together with a single dust opacity index, $\beta$, shared by both components (Appendix~\ref{app:fit}). 
In regions where the 70\,$\mu$m emission falls below the $3\sigma$ level, such as B1 and B2, the warm component was not fitted using the four available wavelengths. Instead, the 70\,$\mu$m flux was treated as an upper limit to constrain the cold component.

We note that the derived parameters remain subject to systematic uncertainties arising from differences in observing conditions, pointing accuracy, beam response, sensitivity, interpolation over positions without spectroscopic coverage, and the imperfect overlap between photometric and spectroscopic bandpasses, the resulting SED fitting should be regarded as a first-order approximation.
Nevertheless, even {accounting for these uncertainties, line contamination significantly biases the inferred dust properties in the B1 and B2 shocked regions (Figure~\ref{fig:sed-point}). Compared with the SED fits using the original bolometric maps without line-contamination correction, the line-subtracted SED fits yield $\rm H_2$ column densities that are two to four times higher, dust temperatures more than 23\% lower, and substantially larger dust opacity indices, $\beta$. Consequently, the corrected $\beta$ values become comparable to those of the protobinary system.}

As noted by \citet{sharma20}, a large-scale filamentary structure extends westward by $\sim6\arcmin$ ($\sim0.6$\,pc) and southward by $\sim5\arcmin$ ($\sim0.5$\,pc) in projection from the protobinary-envelope system, traced by the $J=$1\text{--}0 transitions of $\rm ^{12}CO$, $\rm C^{18}O$, and $\rm N_2H^+$. 
This extended filament is largely filtered out in the original SCUBA-2 850\,$\mu$m map and becomes apparent mainly after combination with \textit{Planck} data (Figure~\ref{fig:jcmt-sma-combination}). 
Although the filament is also detected in the \textit{Herschel} 160\text{--}500\,$\mu$m maps, spectroscopic observations covering its full extent are not available to correct for molecular-line contamination.
 After line subtraction in the regions covered by the JCMT-HARP observations, the residual filamentary emission only remains at the $\sim5\sigma$ level toward the southeast at 850\,$\mu$m and is significantly weaker than the protobinary-envelope system \citep[with an extent of $\sim10^4$\,au;][]{feng22}, the B0-B1 cavity, and the flattened structure extended to 0.3\,pc reported by \citet{bachiller93} and \citet{gueth03}, which could represent the inner dense part of the same large-scale filament. Therefore, we masked the large-scale filamentary component in regions where reliable line-contamination correction is not feasible during the final SED fitting analysis.

The resulting parameter maps are shown in Figure~\ref{fig:sed-map}. The dust temperature exceeds 23\,K toward the dense protobinary system, where the $\rm H_2$ column density reaches $\sim5\times10^{22}\,\mathrm{cm^{-2}}$, corresponding to a mean density of $\sim10^6\,\mathrm{cm^{-3}}$ assuming spherical geometry. 
In contrast, the temperature decreases from $\sim23$\,K at B0 to $\sim13$\,K at B1 and further to below 10\,K at B2 and {across the east–west elongated flattened structure}. 
The $\rm H_2$ column density in the {flattened structure} remains lower than that toward the protobinary system, with typical values of $\rm (1.6$\text{--}$2.0)\times10^{22}\,\rm cm^{-2}$, while the B1 and B2 cavities exhibit significantly lower column densities of $\rm <10^{22}\,cm^{-2}$, consistent with {previous estimates } from $\rm ^{13}CO$ and $\rm C^{18}O$ $J=$1\text{--}0 observations by \citet{umemoto99} and {multitransition CO studies in \citet{lefloch12} and \citet{mendoza14} }.

The dust opacity index map derived from the line-corrected SED fitting shows $\beta\sim$1.7\text{--}1.8 toward both the protobinary system and the B1 cavity, while slightly larger values of $\beta\sim1.8$\text{--}2.3 are found along the B1 cavity wall (Figure~\ref{fig:beta}). 
These values are broadly consistent with those {expected for thermal dust emission}  in the diffuse ISM and dense molecular clouds  \citep[e.g., 1.5\text{--}2.0, ][Planck Collaboration Int. XIV 2014]{schwartz82, ossenkopf94, beckwith90, li01,sadavoy16}. Our results further demonstrate that molecular-line contamination can significantly bias the inferred spectral index. 

 In the first case, if the continuum flux at the long-wavelength end of the SED is systematically overestimated by uncorrected line emission, the SED becomes artificially flattened, leading to an underestimation of $\beta$. Such low values of $\beta$ could potentially mimic the presence of nonthermal emission, such as synchrotron or free-free radiation.

Under an MRN-like grain-size distribution \citep{mathis77}, the dust opacity index, $\beta$, can be used to constrain the maximum grain size \citep[e.g.,][]{miyake93,draine06,birnstiel18}. Variations in $\beta$ from the {flattened structure}  to the outflow cavity may therefore provide insights into the impact of outflows on the grain-size distribution; specifically, we can take a closer look at whether shocks reduce grain sizes through sputtering and/or shattering or instead redistribute larger grains into the envelope \citep{wong16,sabatini24,sabatini25}, an issue that remains under active debate. Our opacity-index map indicates that the maximum grain size is smaller than $\sim100\,\mu$m \citep{miyake93,ossenkopf94}, with grains possibly becoming even smaller toward the shocked regions. This constraint is broadly consistent with grain-size estimates for Class 0/I protoplanetary disks once scattering effects and optical depth are properly taken into account \citep{lij17,agurto19,liu19,silsbee22,maureira25}.

In the second case, the relatively modest values of $\beta$ derived here differ from the  high spectral indices ($\beta+2\sim4.5$--5.5 toward B1--B2) reported from the line-unsubtracted 850\,$\mu$m JCMT and 1.3\,mm MAMBO bolometric data \citep[e.g.,][]{gueth03}. Because the line-to-continuum ratio can differ substantially between 850\,$\mu$m and 1.3\,mm in chemically rich shocked regions, wavelength-dependent line contamination might distort the relative fluxes at the two wavelengths and bias the derived spectral indices toward artificially high values.

We caution, however, that wavelength-dependent line contamination is unlikely to be the sole explanation for the previously reported high spectral indices. 
As discussed by \citet{gueth03}, intrinsic variations in dust properties, grain processing in shocks, and the blending of multiple emission components within the relatively large beams could also contribute to the observed values. Likewise, the current linear resolution of $\sim10^4$\,au limits the robustness of our own grain-size constraints because of the inherent degeneracies among dust opacity, composition, porosity, and temperature gradients along the line of sight.

Disentangling these effects will require high-resolution, multiwavelength continuum observations with accurate line subtraction. Such an analysis, combining higher resolution SMA continuum observations with line-subtracted 0.85 and 1.3\,mm data, will be presented in a forthcoming study.
}

\begin{figure*}
  \centering
        \begin{tabular}{lccc}
\rotatebox{90}{\parbox{5cm}{\centering $\Delta$ Dec.($\arcsec$) }}
&\includegraphics[clip, trim=0.0cm 0.0cm 0.01cm 0.0cm,height=7.5cm]{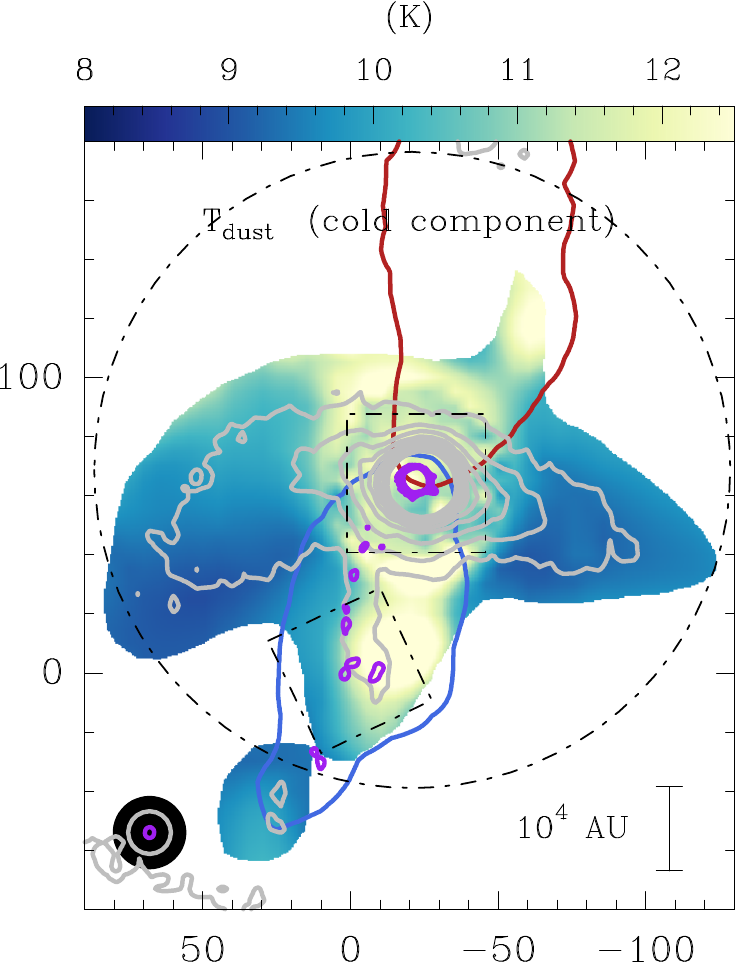}
&\includegraphics[clip, trim=1.2cm 0.0cm 0.01cm 0.0cm,height=7.5cm]{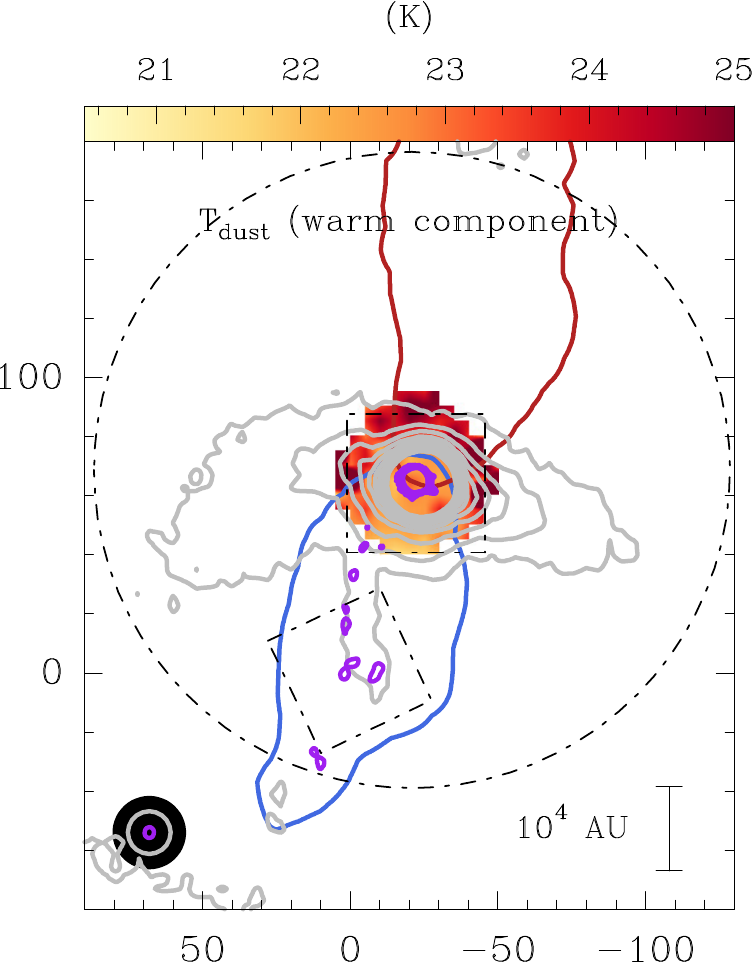}
&\includegraphics[clip, trim=1.2cm 0.0cm 0.0cm 0.0cm,height=7.5cm]{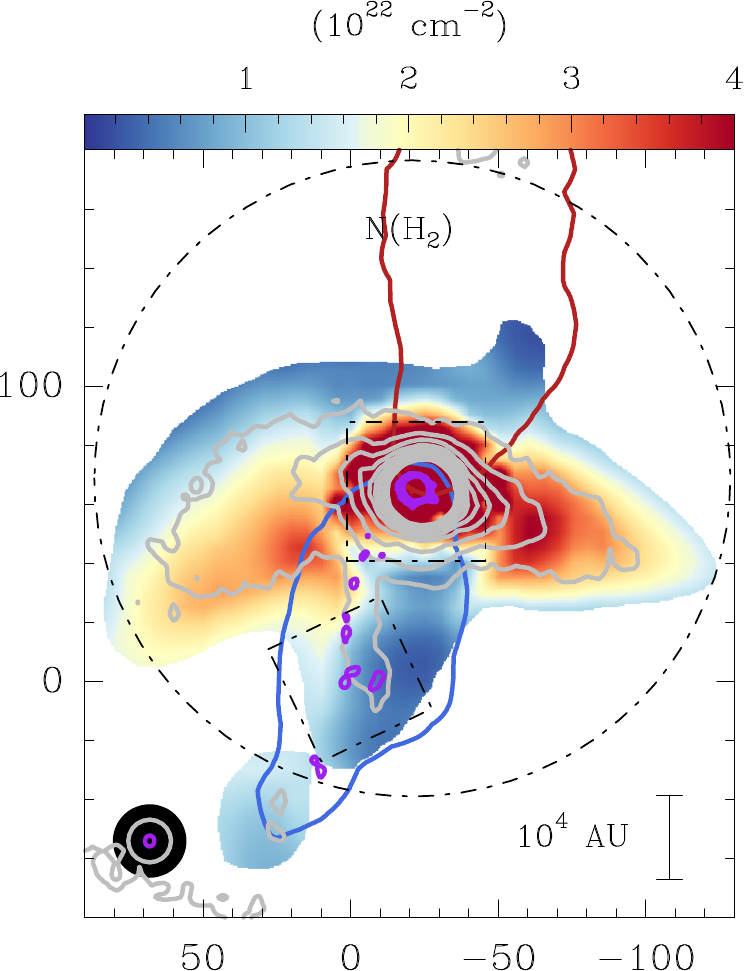}\\

& &$\Delta$R.A.($\arcsec$) &
\end{tabular}
\caption{
Dust temperature and $\rm H_2$ column density maps derived from pixel-by-pixel SED fitting to the line-contamination-corrected \textit{Herschel} (70--350\,$\mu$m) and JCMT-\textit{Planck} (850\,$\mu$m) continuum data.
The fitting adopts  a two-component model and assumes a gas-to-dust mass ratio of 100. 
Gray contours show the  line-contamination-corrected  850\,$\mu$m continuum emission from the JCMT-\textit{Planck} combination,
while the red- and blue-shifted outflow lobes traced by JCMT CO\,(3\text{--}2) emission are shown in red and blue contours, respectively.
Purple contours represent the SMA 1.3\,mm dust continuum derived from line-free spectral channels. Contour levels are identical to those in Figure~\ref{fig:source}.
The gray circle in the lower left corner of each panel denotes the JCMT beam (14.5\arcsec), the purple ellipse marks the SMA synthesized beam ($\rm \sim3.4\arcsec$), and the black filled circle indicates the common angular resolution (24.9\arcsec) used for the parameter maps. 
The black dashed circle indicates the region where spectroscopic products from the COPS survey  \citep{yang18} are used to remove line contamination from the 250\text{--}350\,$\mu$m bolometric maps. 
The black dashed boxes mark the protobinary-envelope region covered by the CDF survey \citep{green16} and the B1 shocked region covered by the CHESS survey \citep{codella10,lefloch10}, where spectroscopic products are used to subtract line emission from the 70 and 160\,$\mu$m bolometric maps.
Regions where the SED fitting uncertainties exceed 50\% of the derived values, and where the smoothed 850\,$\mu$m emission falls below the $\rm 3\,\sigma$ level, are masked.
}\label{fig:sed-map}
\end{figure*}

\begin{figure*}
  \centering
      \begin{tabular}{lccc}
\rotatebox{90}{\parbox{10cm}{\centering $\Delta$ Dec.($\arcsec$) }}
&\includegraphics[clip, trim=0.0cm 0.0cm 0.02cm 0.0cm,height=10cm]{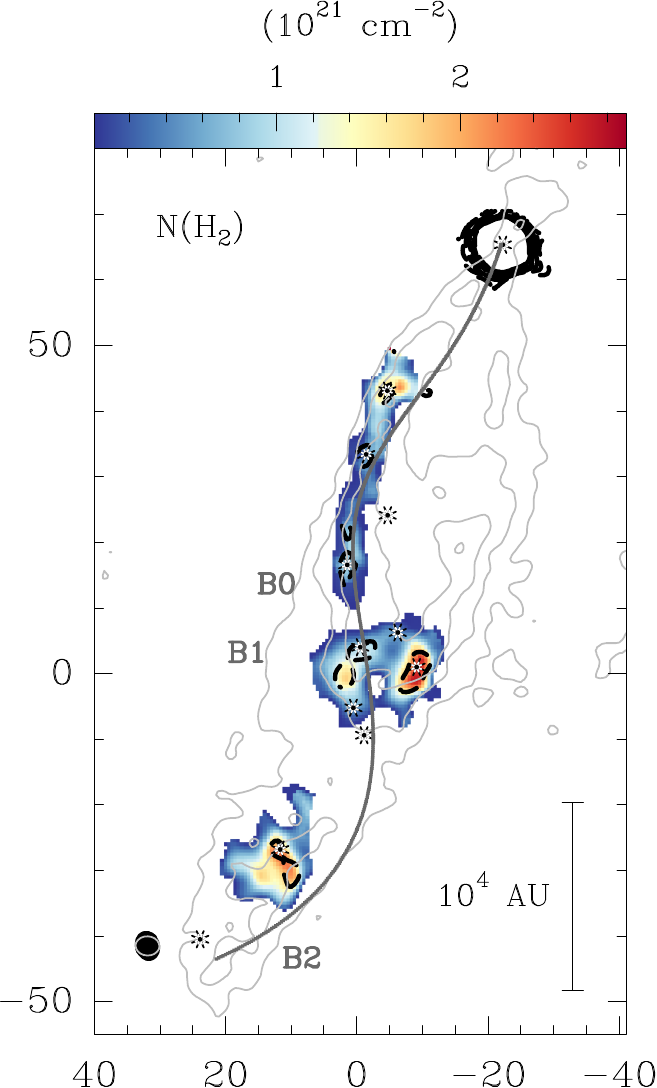}
&\includegraphics[clip, trim=1.5cm 0.0cm 0.0cm 0.0cm,height=10cm]{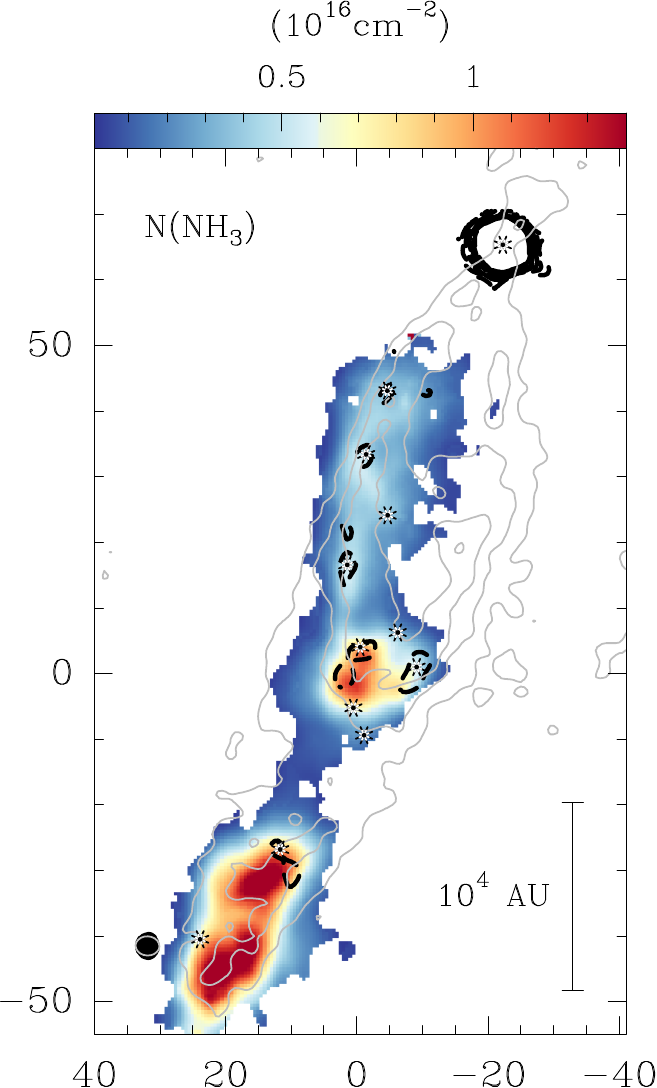}
&\includegraphics[clip, trim=1.5cm 0.0cm 0.0cm 0.0cm,height=10cm]{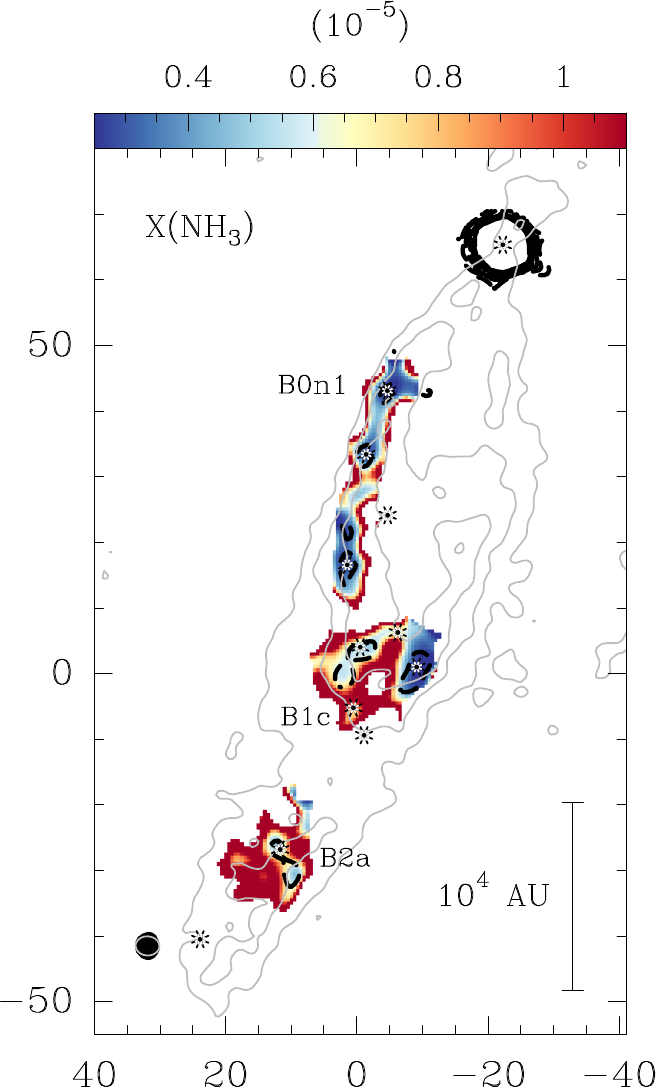}\\

& &$\Delta$R.A.($\arcsec$) &
\end{tabular}

\caption{
$Left$: $\rm H_2$ column density map derived from the 1.3\,mm dust continuum observed with the SMA, assuming a gas-to-dust mass ratio of 100, optically thin dust, and adopting the dust temperature from the gas kinetic temperature map traced by the $\rm NH_3$ (1,1)\text{--}(6,6) transitions \citep{feng22}. Continuum with emission $\rm <3\sigma$ is masked.
$Middle$: $\rm NH_3$ column density map {toward the shocked region} obtained from LVG modeling of the $\rm NH_3$ (1,1)\text{--}(6,6) transitions, combining both ortho and para species, adopted from \citet{feng22}. {Both column densities toward the protobinary system are not reported because the line profiles exhibit multiple blended velocity components, strong optical-depth effects, and significant spatial overlap between the protobinary, disk, and envelope at the available angular resolution, preventing a reliable single-component analysis.}
$Right$: Map of the relative abundance ratio of $\rm NH_3$ with respect to $\rm H_2$. 
{Black dashed contours  and black  markers of the representative positions} in each panel is the same shown  in Figure~\ref{fig:source}. 
{The gray contours, starting from 5$\sigma$ ($\sigma = 0.46$ Jy beam$^{-1}$) and increasing with a step of 5$\sigma$, show the CO \,(1\text{--}0) emission}, obtained from the NOEMA and IRAM-30\,m combination \citep{gueth96}.
The dark gray solid curve in the left panel traces the precessing jet path identified by \citet{podio16}. 
The black filled ellipse and gray circle  in the lower left corner of each panel denote the SMA synthesized beam at 1.3\,mm  and the beam of CO\,(1\text{--}0) emission, respectively.
}
\label{fig:nh2-sma}
\end{figure*}

Nevertheless, even a single-band continuum observation at a linear resolution of $\sim10^3$\,au provides valuable insight. Zooming in on this linear resolution, we derived an $\rm H_2$ column density map from the 1.3\,mm continuum using the standard single-wavelength dust-emission formalism of \citet{schuller09}, assuming optically thin emission, a gas-to-dust mass ratio of 100, and thermal coupling between the dust and gas along the cavity walls. We adopted the gas kinetic temperature map derived from $\rm NH_3$ (1,1)\text{--}(6,6) observations at a comparable angular resolution \citep{feng22}, which shows temperatures of $\sim$80\,K in the cavity to $\sim$120\,K along the cavity walls. 
As shown in  Figure~\ref{fig:nh2-sma} (left panel), the inferred column densities range from the $\rm 3\sigma$ sensitivity limit of $\rm 1.5\times10^{20}\,cm^{-2}$ in the low-density cavity to  (1\text{--}3)$\rm \times10^{21}\,cm^{-2}$ in the dense cavity wall. 
Assuming that each dusty clump is spherical, the corresponding volume densities are  (2\text{--}5)$\rm \rm \times10^5\,cm^{-3}$, in good agreement with those inferred from large-velocity-gradient (LVG) modeling \citep{gomezruiz13,gomezruiz15,feng20a}. 
The clump masses are in the range of (0.4\text{--}3)$\rm \times 10^{-3}\,M_\odot$, with  surface densities in the range of (0.6\text{--}1.2)$\rm \times 10^{-2}\,g\,cm^{-2}$. 
We tested dust-opacity indices of $\beta=1.5$, 1.8, and 2.3 and found that they produce only minor variations in the adopted opacity, $\kappa_\nu \sim 0.899(\nu/230\,\rm{GHz})^\beta$, appropriate for grains with thin ice mantles at densities of $\rm 10^5\text{--}10^6$\,$\rm {cm^{-3}}$. 
Moderate temperature variations over 80\text{--}130\,K  are therefore unlikely to account for the observed intensity variations, which are instead dominated by changes in $\rm H_2$ column density. For example, B1b and B1c share similar kinetic temperatures  (80\,K)  but differ in the emission at 1.3\,mm by a factor of at least 3.

Because of the missing short-spacing information in the SMA observations, the degree of spatial filtering is uncertain and likely varies across the mosaic. We therefore excluded the 1.3\,mm SMA continuum data from the multiband SED fitting. Nevertheless, a comparison with the SED model indicates that the SMA recovers nearly all of the total 1.3\,mm continuum flux toward the central protobinary system (mm), but only 15\%--46\% toward the B0, B1, and B2 cavity walls.

Although correcting the SMA-derived $\rm H_2$ column densities at a linear scale of $\sim$1200\,au for the missing flux (by factors of 2\text{--}7, based on the measured flux recovery fraction) brings them into agreement with those inferred from lower angular-resolution ($\sim10^4$\,au) SED analysis, the native-resolution SMA observations show no corresponding increase in column density toward the inner outflow cavity. Instead, the column density remains comparable to, or even lower than, that of the surrounding colder filament. 
 This behavior contrasts with the monotonic density increase toward smaller radii predicted by protostellar collapse models and commonly observed in quiescent protostellar envelopes \citep[e.g.,][]{larson69,shu77,shirley02}. 
 Such expectations, however, may not apply in the outflow-envelope interaction zone, where shocks, gas removal, and dust processing can substantially modify the local density structure. We note that in the protostellar shocked region with {intermediate} densities ($\rm 10^5\text{--}10^6\,cm^{-3}$), dust–gas thermal coupling might be inefficient, depending on the shock type \citep[e.g., ][]{draine80,hollenbach89,draine93}. 
As a result, the dust temperature in L1157 is likely lower than the gas temperature. 
{Since the dust emission at mm wavelength is highly sensitive to temperature below $\sim$50\,K, uncertainties in the dust temperature can introduce substantial systematic errors in the derived column densities.}
Under this scenario, the apparent {decline} in dust emission, from the $5\text{--}7\sigma$ peak (e.g., B1b) to marginal ($\sim3\sigma$, B1c) detections and all the way down to no detections ($<3\sigma$, B1i), does not necessarily {imply a corresponding decrease} in the dust column density. Instead, it may reflect enhanced radiative heating at the cavity walls, where dust illuminated by the central protostar appears brighter at mm wavelengths. Higher resolution observations at shorter wavelengths, {probing spatial scales of $\sim \rm10
^3$\, au} will be required to disentangle the {relative roles of} dust temperature, column density, grain properties, and possible nonthermal emission  in shaping the observed dust emission variations ({Yang et. al. in prep.}).

\subsection{Shock induced dust evolution and  chemistry}\label{sec:abundnace}

Although a rich inventory of molecular lines has been detected toward the L1157 B1 and B2 shocks, {direct  dust continuum constraints on the shocked cavity structures remained limited in previous studies.}
Consequently, chemical interpretations have primarily relied on the spatial distributions and relative abundances of molecular tracers. 
In particular, species distributions {observed at a linear resolution of $\rm 10^3$\,au} have often been compared with that of HDCO, which is thought to form predominantly on grain mantles and to be released into the gas phase \citep{fontani14}, to distinguish between gas-phase and grain-surface formation pathways \citep[e.g.,][]{codella15,codella17,lefloch24}. 
Formation routes were typically constrained by reproducing the observed relative molecular abundance ratios along the outflow, with projected distance from the protostar adopted as a proxy for chemical timescale.

The dust morphology revealed by our observations now provides {direct observational constraints} on the distribution and survival of dust within these shocked regions, enabling a more explicit assessment of the assumptions underlying such chemical modeling. 
{Comparison with the gas kinetic temperature map \citep{feng22} shows that the detected dusty clumps reside in regions} where the gas temperature exceeds 80\,K, likely due to heating by the precessing jet. 
However, the spatial offset between the dust continuum and gaseous NH$_3$ emission suggests that molecular spatial distributions alone may not fully {trace the underlying dust distribution and} dust-gas interactions in shocked environments.

Although $\rm NH_3$ lines  are widely used as gas thermometer  in star-forming regions, whether this species is predominantly formed from grain-surface or gas-phase remains under debate  \citep{tielens21}.
 In cold ($\sim$10\,K), dense ($n\gtrsim10^4$\,cm$^{-3}$) environments \citep{bergin07}, 
recent chemical modeling shows that $\rm NH_3$ can form efficiently through gas-phase reactions once a significant fraction of nitrogen becomes molecular, through slightly endothermic hydrogen abstraction reactions followed by dissociative recombination of $\rm NH_4^+$ \citep[e.g., ][]{hilyblant10, sipila19}. 
 However, the resulting gas-phase abundance typically remains below $\sim 10^{-7}$.
In contrast, $\rm NH_3$ is expected to form predominantly on grain surfaces via successive hydrogenation of atomic nitrogen and to be released into the gas phase through thermal or nonthermal desorption.
Gas-phase production of $\rm NH_3$ at abundances as high as  $\sim 10^{-5}$ in L1157 B1 has been predicted only under extreme conditions, namely: temperatures $\gtrsim800$\,K and densities $\rm \gtrsim10^5\,cm^{-3}$ sustained for $\gtrsim10^4$\,yr without significant freeze-out \citep{viti11}. The present observations, combining high spatial resolution with time-dependent chemical modeling, offer a unique opportunity to test whether protostellar shocks can indeed generate and maintain such conditions.

At a linear resolution of $\rm 10^3$\,au,  the optical-depth–corrected $\rm NH_3$ column density map shows an enhancement toward the U-shape head of bow-shock  B1 and B2 compared with the  cavity wall (B0) by a factor of 3, reaching $\rm 10^{16}\,cm^{-2}$ (Fig.~\ref{fig:nh2-sma}, middle panel,  see \citealp{feng22}). This value is higher than those reported by  {\citet{bachiller93,tafalla95,umemoto99} by a factor of 20\text{--}100,  primarily because those studies either relied on single-dish observations, assumed the  $\rm NH_3$ emission to be optically thin, or did not account for both the ortho- and para-$\rm NH_3$ populations when deriving the total NH$_3$ column density.}  Using the $\rm H_2$ column density map derived under the assumptions of optically thin dust and gas–dust thermal coupling (Section~\ref{sec:profile}), we estimated the relative $\rm NH_3$ abundance with respect to $\rm H_2$. The abundance increases from the dusty peak near B0 ($\sim2\times10^{-6}$) by at least a factor of 4 toward the B1 shock front and by at least a factor of 2 toward B2. It reaches {$\gtrsim10^{-5}$} at positions such as B1c and B2a, where the $\rm H_2$ column density is constrained only by 3\text{--}5$\sigma$ upper limits.
Even if we are accounting for a possible underestimation of  $\rm H_2$ column density due to optically thin dust emission or for overestimated and spatially varying dust temperatures (as discussed in Section~\ref{sec:profile}), the trend of increasing $\rm NH_3$ abundance from B0 to B1 and B2 by a factor of $\rm \sim3$ remains robust.
The enhanced $\rm NH_3$ abundances in the shock regions, particularly toward the B1 shock front and B2 ($\rm >10^{-6}$ assuming dust temperatures  $\gtrsim50$\,K), are significantly higher than the typical values observed in star-forming regions \citep[$\sim10^{-9}$\text{--}$10^{-7}$;][]{friesen09, herbst09, caselli12b, jorgensen20}.

To understand the high abundance of $\rm NH_3$ and its enhancement  from the cavity wall to the shock head, we constructed a coupled {physicochemical} model describing the evolution of gas property toward L1157 B0, B1, and B2 regions (see Appendix~\ref{app:model} for details). Given that the {L1157 outflow cavity} is shaped by a precessing jet with an inclination of $\sim73^\circ$ and a precession angle of $\sim8^\circ$ \citep{podio16}, we adopt a simplified geometry in which the outflow is continuously launched from the protobinary system. The B0, B1, and B2 regions observed today are assumed to be impacted by shocks {produced by} gas parcels ejected at different epochs. Assuming that these parcels reach B0, B1, and B2 at approximately the same observational epoch, the projected separations imply ejection intervals of 500\text{--}800\,yr between B0 and B1 and 700\text{--}1000\,yr between B1 and B2. Owing to jet precession, each region is assumed to experience {an independent} shock event. {Consequently, the gas currently driving the B2 shock follows a different trajectory and is not assumed to have propagated through the present-day B0 or B1 shock positions at earlier times.}

The {physicochemical} evolution at each position is modeled in two stages: a pre-shock phase and a post-shock phase. During the pre-shock phase, all regions are assumed to share envelope-like conditions, with a hydrogen nuclei density of $n_{\text{H}}$ = 1$\times$ 10$^5$ $\text{cm}^{-3}$ and a gas temperature of 10\,K. 
After  $\rm \sim10^6$\,yr of chemical evolution, the resulting abundances are adopted as initial conditions for the shock phase.
In the outflow-shock phase, gas parcels are ejected sequentially from the protostar. We tracked three representative parcels that reach the positions B0n1, B1c, and B2a after 900\,yr, 1750\,yr, and 2500\,yr , respectively, triggering shocks at those locations. Regardless of small differences in arrival times, the observed chemistry is assumed to reflect post-shock conditions at all three positions.
 Although different shock types have been explored in previous studies, {the resulting NH$_3$ abundances do not differ substantially within the parameter space relevant to L1157.
  We therefore adopted C-type shock models for all regions, consistent with previous studies of the source \citep{viti11,holdship17}.  Following \citet{lefloch21}, the shock velocity is assumed to decrease monotonically along the flow, from 60\,$\rm km\,s^{-1}$ at B0n1 to 40\,$\rm km\,s^{-1}$ at B1c and 20\,$\rm km\,s^{-1}$ at B2a.}

The shock converts kinetic energy into thermal energy, {raising} the gas temperature to several thousand kelvin before cooling to $\gtrsim100$\,K, $\sim110$\,K, and $\lesssim80$\,K at B0n1, B1c, and B2a, respectively \citep{feng22}. 
{While these temperatures vary among locations to reflect the global shock gradients, they are assumed to remain stationary for any given shock layer during its subsequent chemical evolution.} 
Figure~\ref{fig:physicalchanges} illustrates the temporal evolution of the gas properties at each location following the propagation of each parcel ejected from the protostar. 
The corresponding post-shock hydrogen nuclei densities are $1\times10^{6}$, $7\times10^{5}$, and $\rm 4\times10^{5}\,cm^{-3}$, respectively. 
Dust and gas are assumed to be thermally decoupled, with dust heated from  10 to 50\,K {after} the passage of the shock. 
{Following \citet{wakelam21}, we adopt the cosmic-ray ionization rate (CRIR) prescription of \citet{neufeld17} as an input parameter rather than deriving it through chemical fitting. The resulting time-dependent CRIR varies between (1\text{--}5)$\rm \times10^{-17}\,s^{-1}$ over the course of the simulation.}

\begin{figure}
\includegraphics[width=0.5\textwidth]{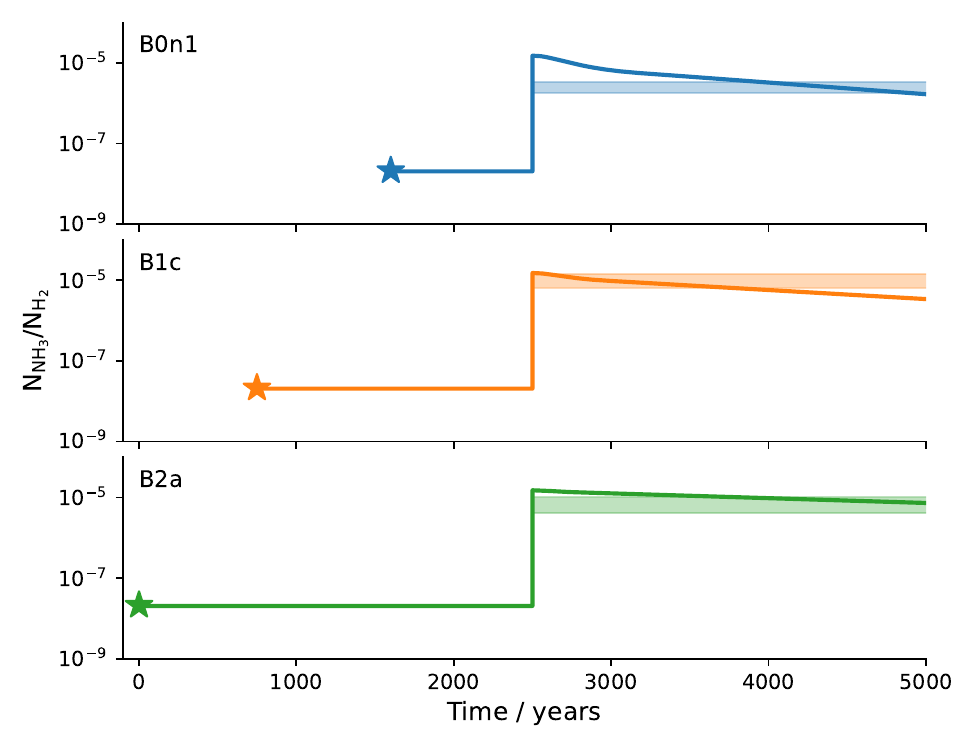}
\caption{
Temporal evolution of the gas-phase $\rm NH_3$ abundance from the ``three-phase" chemical model, compared with the observed relative abundance ratios of $\rm NH_3$ with {respect} to $\rm H_2$  toward the B0n1, B1c, and B2a regions.
The model adopts a time-dependent {CRIR} following \citet{wakelam21} and assumes a gas-to-dust mass ratio of 100.
The shaded regions indicate the observed abundance ranges; the upper and lower bounds reflect uncertainties associated with assuming gas–dust thermal coupling (i.e., equal gas and dust temperatures) and a dust temperature of 50\,K.
Star symbols mark the ejection times of individual parcels from the protostar. Gas-phase $\rm NH_3$ abundances are not shown prior to each parcel ejection.
}\label{fig:gasNH3_abun}
\end{figure}

Chemical evolution is modeled using the ``three-phase" astrochemical code \textsc{Nautilus} \citep{ruaud16,wakelam24}, which self-consistently includes gas-phase chemistry, grain-surface and ice-mantle reactions, adsorption, thermal and nonthermal desorption, and shock-induced sputtering.
Prior to shock arrival, gas-phase $\rm NH_3$ produced mainly via dissociative recombination of NH$_4^+$ {remains} below $\rm 10^{-7}$. 
During the shock phase, although elevated temperatures enhance gas-phase reaction rates, pure gas-phase chemistry, even under an extreme {CRIR} of ($\rm 10^{-15}\,s^{-1}$) , fails to produce gaseous 
$\rm NH_3$ abundances over $10^{-6}$ (Figure~\ref{fig:gasNH3_diff_CR}).
In contrast, $\rm NH_3$ forms efficiently on dust grains (denoted by the $J$ hereafter) through successive hydrogenation reactions ($J$NH $\rightarrow$ $J\rm NH_2$ $\rightarrow$ $J\rm NH_3$), reaching grain-surface abundances of $\sim10^{-5}$ prior to shock arrival. Shock-induced nonthermal desorption processes, including sputtering and grain–grain collisions \citep{caselli97}, rapidly release this reservoir into the gas phase, producing the sharp gaseous $\rm NH_3$ {abundance} enhancements shown in Figure~\ref{fig:gasNH3_abun}.

Following shock passage, the abundance of gaseous $\rm NH_3$  gradually declines as it re-adsorbs onto surviving dust grains during the cooling phase. While the exact post-shock duration at each position is unconstrained from the previous observations, the shock velocities at B1c and B2a indicating from observations suggest that significant dust destruction is unlikely. Instead, the elapsed time since shock passage at these locations might be shorter than at B0n1. Consequently, insufficient time for substantial $\rm NH_3$ freeze-out likely accounts for the higher gaseous abundances observed toward the B1 shock head and B2.

The simplified {physicochemical} framework adopted in this study provides a  first-order interpretation of the observed $\rm NH_3$ enhancement, though several physical complexities need further investigation. 
First, the assumption of {spatially varying but locally} constant post-shock gas temperature and density may not fully capture the time-dependent thermal and dynamical evolution of shocked material. 
Second, if regions such as B0n1 and B1c have experienced multiple shock passages, our current model cannot rule out the possibility that $\rm NH_3$ (or its precursors) was released at earlier locations and subsequently advected downstream to B2a, where it may have continued to chemically interact with the ambient gas during transport. Such cumulative or multishock processing is not explicitly included. 
Third, while our results strongly {suggest} that the high $\rm NH_3$ abundances ($\rm 10^{-5}$) are best explained by grain-surface formation followed by shock-induced desorption into the gas phase, the origin and evolution of the dust grains toward the outflow cavity wall themselves remain uncertain. 
Our model cannot distinguish whether the grains formed in the immediate protostellar envelope and were subsequently entrained by the outflow, or whether they originate from redistributed ambient material as the outflow excavates a low-density cavity, sweeping material outward and concentrating it in the cavity walls.
Finally, the presence of strong $\rm NH_3$ line emission in regions with marginal or undetected  dust continuum emission {at mm wavelength} might reflect either low dust temperatures or low dust column densities. With observations currently limited to mm wavelengths, grain sizes and dust masses remain poorly constrained. We therefore do not attempt to quantify how efficiently outflow shocks modify the grain population or which physical mechanisms dominate.
Potential processes include grain–grain collisions leading to shattering or vaporization \citep[e.g.,][]{jones94,jones96,guillet09,guillet11}, sputtering \citep[e.g.,][]{draine79,caselli97}, grain charging, and magnetic decoupling \citep[e.g.,][]{draine80,pilipp94,guillet07}, and ice-mantle stripping \citep[e.g.,][]{jimenez08,podio14}. 
Detailed chemical modeling is beyond the scope of this work. Instead, we highlight the key observational constraints that will guide future, more comprehensive shock–chemistry models.

\section{Conclusions}\label{sec:conclusion}

Our new JCMT 825\text{--}906\,$\mu$m and SMA 1.1\text{--}1.4\,mm observations provide {high-sensitivity, spatially resolved images of the thermal dust emission in the} L1157 B0–B1–B2 region, an archetypal chemically rich protostellar outflow {shaped} by multiple shocks, spanning spatial scales from {0.4}\,pc down to 1200\,au. By {quantitatively} removing line contamination from the broadband 70\text{--}850\,$\mu$m bolometric maps, we were able to recover the dust temperature structure not only in the warm shocked gas, {as well as the extended protobinary envelope}, revealing a coherent dust thermal connection between the outflow cavities and the surrounding environment.

At a linear resolution of $\rm 10^4$\,au, the dust opacity index {shows a slight decrease from the B1 cavity wall ($\sim $2.3) to the cavity and the central protobinary system ($\sim $1.8)}, indicating  that dust grains have not yet reached mm sizes in these shocked regions.
Compared to previous bolometric analyses, our two-component SED fitting further demonstrates the  importance of {quantitatively} correcting for line contamination in chemically active environments.

At a linear resolution of $\sim10^3$ au, the dust emission is resolved into compact dusty structures {distributed} along the precessing jet, {identifying} the sites where molecules such as $\rm NH_3$ can form and be processed. The relative abundance of $\rm NH_3$ with respect to $\rm H_2$ is strongly enhanced in these regions, exceeding $\sim10^{-5}$ and reaching peak values at the B1 shock front, even where the 0.85 and 1.3\,mm dust continuum emission falls below the $\rm 5\sigma$ detection limit. Our chemical modeling, {starting with a benchmark gas-phase analysis, demonstrates that gas-phase reactions alone cannot reproduce the observed abundances, even under extremely elevated {CRIR}s. This provides empirical evidence that}  $\rm NH_3$ must form predominantly on grain surfaces and be rapidly released into the gas phase through shock-induced sputtering. Although {the present observations cannot uniquely determine whether the lack of detectable dust emission toward the B1 shock front is caused by low dust temperatures, reduced dust column densities, or changes in grain properties}, our modeling suggests that the enhanced $\rm NH_3$ abundance in that case could result from insufficient re-adsorption (freeze-out) of gaseous $\rm NH_3$ onto dust grains.

Taken together, these results {reveal a close interplay between} dust evolution and molecular chemistry in protostellar shocks through a cyclic process of grain processing, mantle release, and reformation. These results also caution against inferring molecular formation pathways solely from the spatial distributions of gaseous species, highlighting the essential role of direct {imaging of the dust distribution} in chemically rich shocked regions.

\begin{acknowledgements}

S.F. acknowledges support from the National Key R\&D program of China grant (2025YFE0108200) and National Science Foundation of China (12373023,  12133008).

H.B.L. is supported by the National Science and Technology Council (NSTC) of Taiwan (113-2112-M-110-022-MY3).

Y.L, X.J, and D.Q are supported by Key R\&D Program of Zhejiang grant (2026SSYS0001), the Leading Innovation and Entrepreneurship Team of Zhejiang Province of China (2023R01008) and National Science Foundation of China (12373026).

Z.-Y. Z. acknowledges the support of the National Natural Science Foundation of China (12533003,1257030642).

F.Du is supported by National Key R\&D program of China grant (2023YFA1608004) and National SKA Program of China (2025SKA0140100).

{The Starlink software \citep{currie14} is currently supported by the East Asian Observatory.}

This research made use of NASA's Astrophysics Data System.

 \end{acknowledgements}

\bibliographystyle{aa}          
%\bibliography{/Users/siyifeng/GoogleDrive/HMSFR.bib}
\bibliography{aa59452-26.bib}

\setcounter{section}{0}
\renewcommand{\thetable}{A\arabic{section}}
\setcounter{table}{0}
\renewcommand{\thetable}{A\arabic{table}}
\setcounter{figure}{0}
\renewcommand{\thefigure}{A\arabic{figure}}
\begin{appendix}

\section{Line detection and continuum subtraction with JCMT-HARP}\label{app:linejcmt}
To ensure the accuracy of the 0.85\,mm bolometric continuum, we performed a systematic subtraction of line emission within the JCMT-HARP frequency coverage (329.9\text{--}362.4\,GHz). This procedure is essential for isolating the thermal dust continuum contribution in the chemically rich L1157 shocked environment.

Figure~\ref{fig:line1} presents beam-averaged spectra extracted from three representative regions, the protobinary system (mm), and the shocked regions B1a and B2a. 
To illustrate the spectral complexity relevant to the continuum subtraction, we label prominent lines detected above the $5\sigma$ level and spatially extended beyond the JCMT beam size ($14.5\arcsec$).

Because the primary goal of the present work is the continuum analysis, the line identifications provided here should be regarded as representative rather than exhaustive. Candidate transitions were selected using the CDMS \citep{muller05} and JPL \citep{pickett98} molecular databases by comparing rest frequencies, Einstein coefficients ($A_{ij}$), and upper level energies ($E_u/k_B$) expected under protostellar shock conditions. We prioritize the most probable identifications based on their spectral intensities and physical plausibility.
A comprehensive census of the molecular-line forest, including detailed line identification, kinematic analysis, column-density determination, and chemical modeling, will be presented in a dedicated follow-up study.

\begin{figure*}[!t]
  \centering
\includegraphics[width=1.0\textwidth]{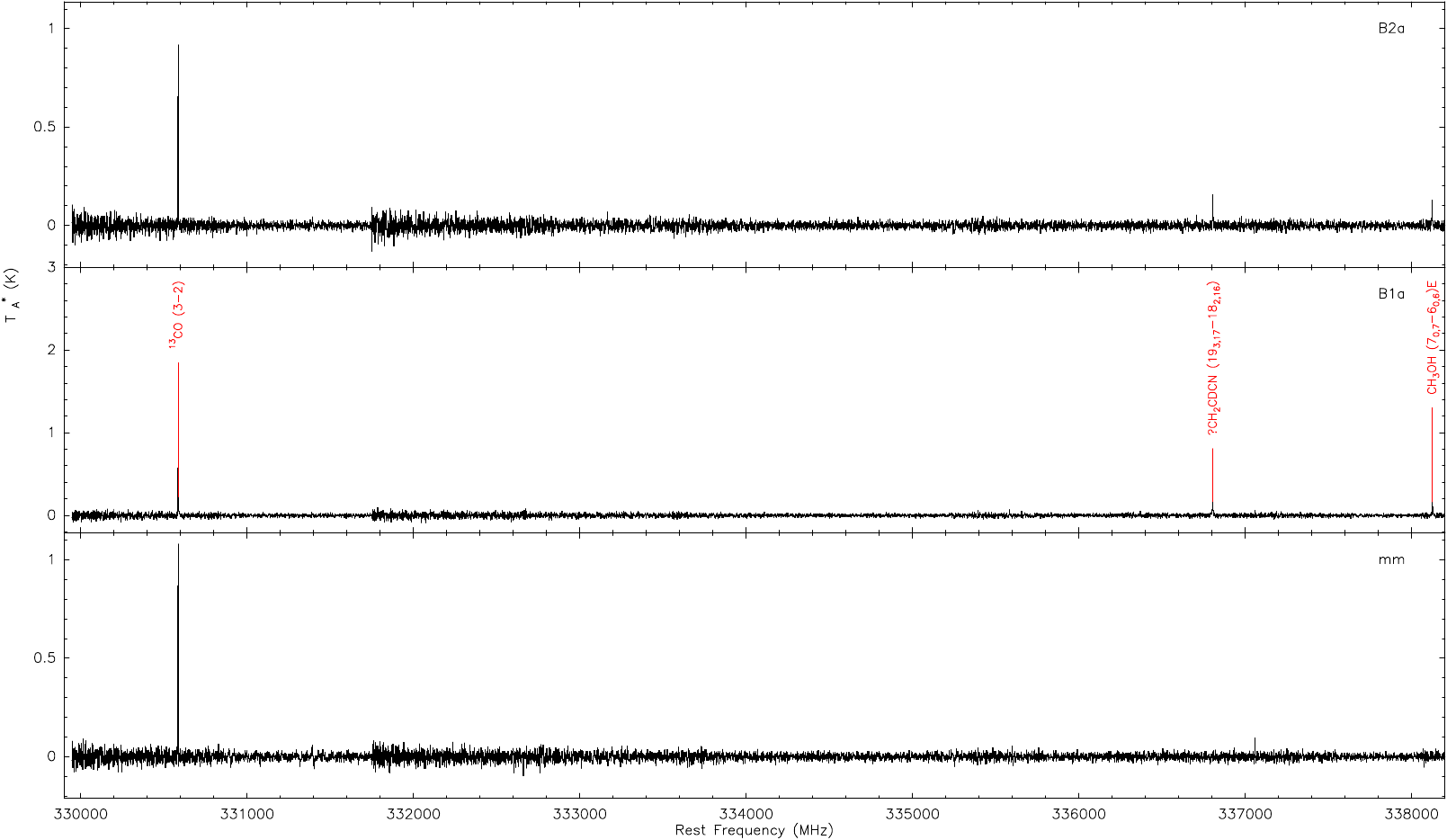}
\includegraphics[width=1.0\textwidth]{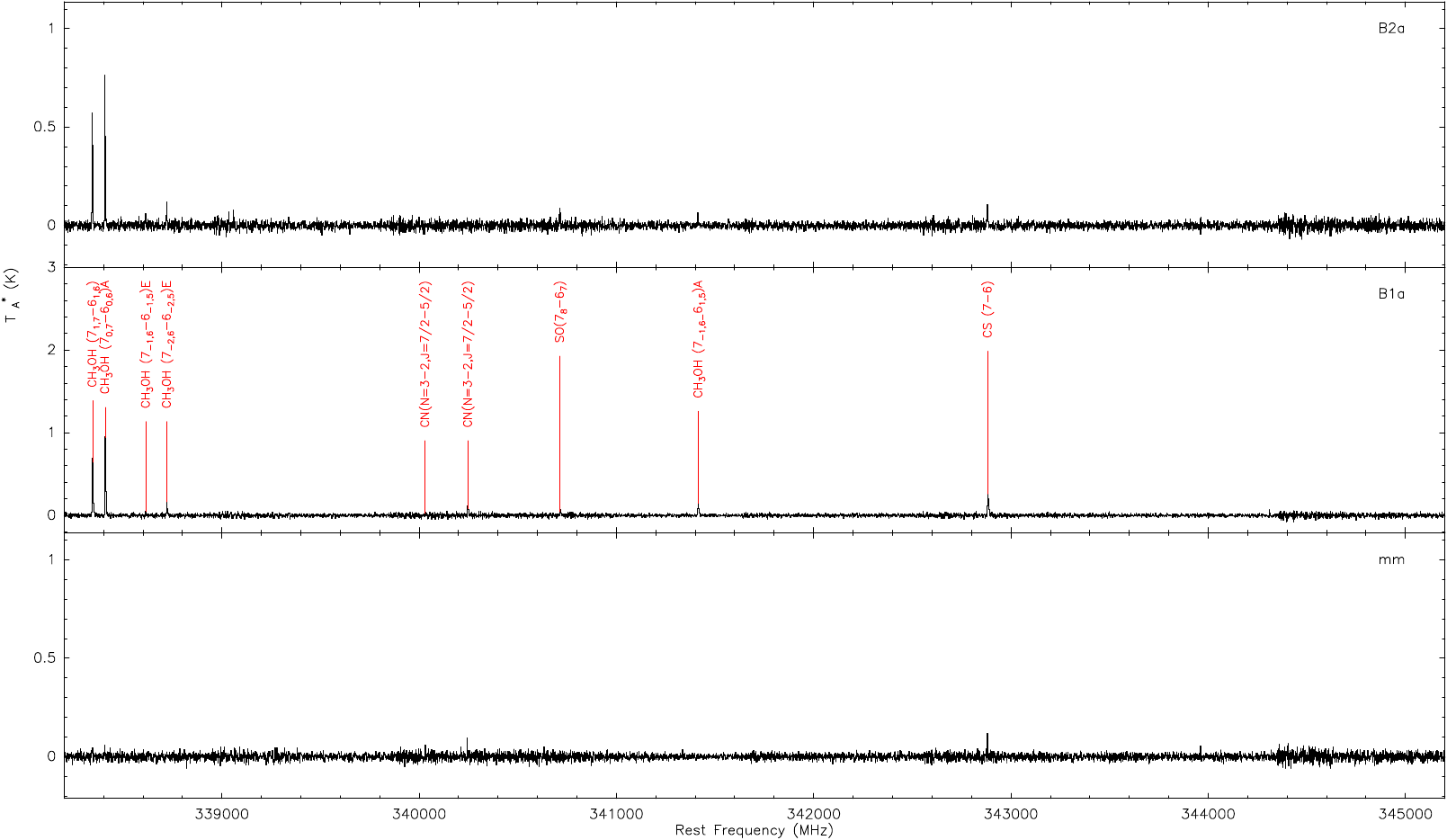} 
\caption{
Beam-averaged spectra extracted from the L1157 protobinary envelope (mm) and the shocked regions B1a and B2a using JCMT-HARP over the frequency range 329.9\text{--}362.4\,GHz. 
Labeled transitions correspond to emission detected above the $5\sigma$ level and spatially extended beyond the $14.5\arcsec$ JCMT beam. 
The identified lines are primarily shown to illustrate the spectral complexity of the region and to support the molecular-line subtraction applied to the bolometric continuum maps.
}
\label{fig:line1}
\end{figure*}

\addtocounter{figure}{-1}
\begin{figure*}[!t]
  \centering
\includegraphics[width=1.0\textwidth]{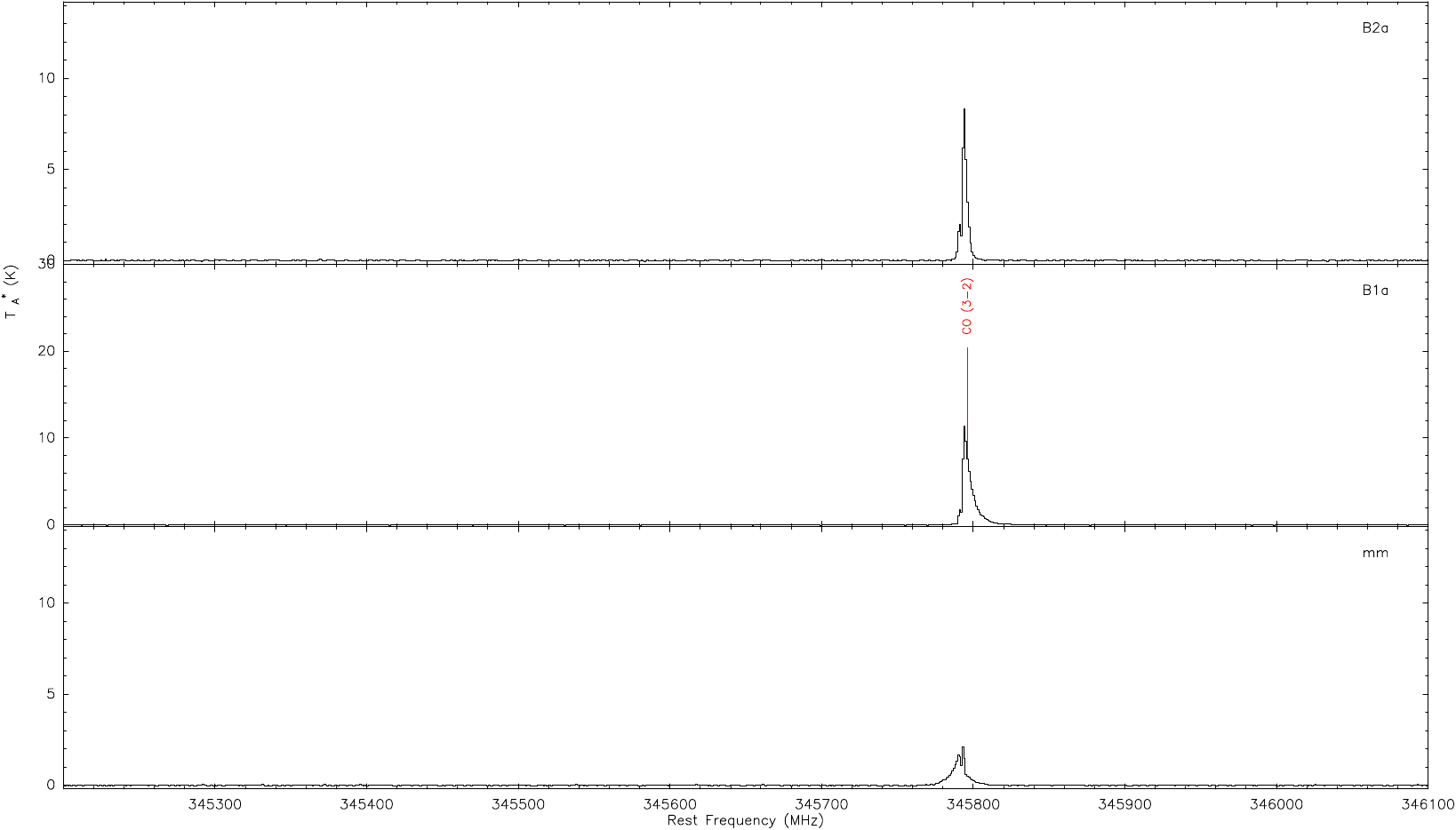}
\includegraphics[width=1.0\textwidth]{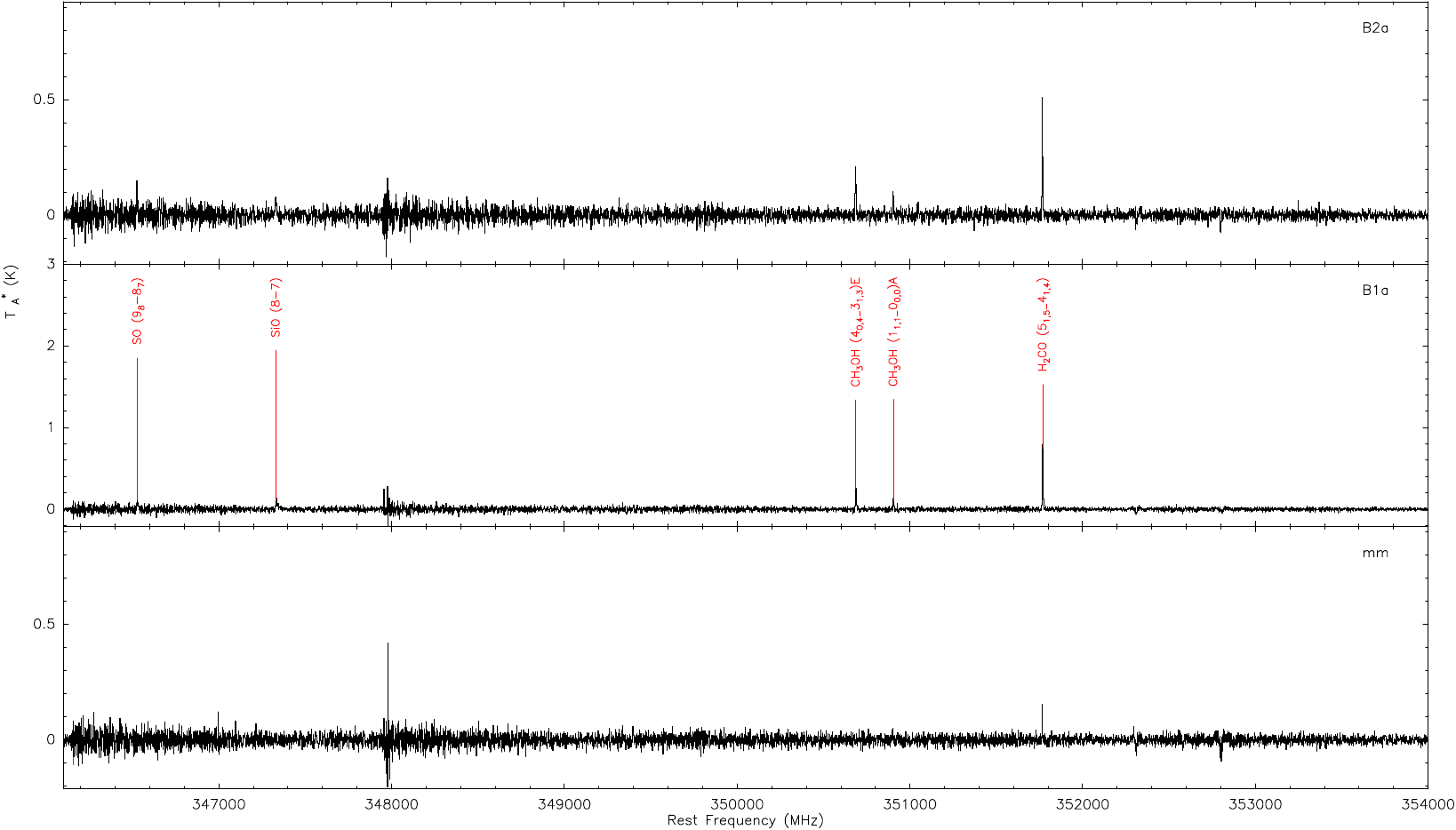} 
\caption{
Continued
}
\label{fig:line2}
\end{figure*}

\addtocounter{figure}{-1}
\begin{figure*}[!t]
  \centering
\includegraphics[width=1.0\textwidth]{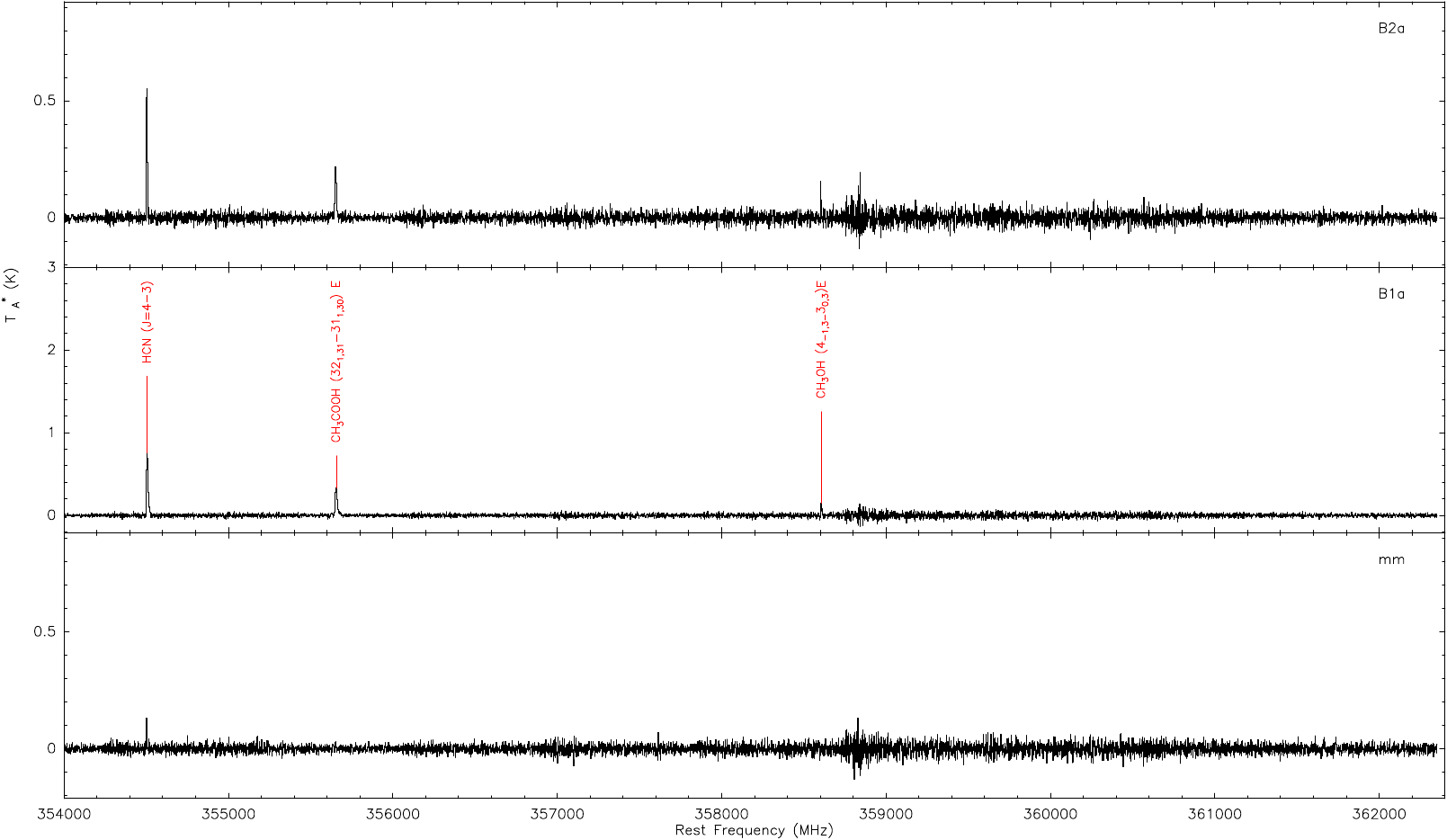}
\caption{
Continued
}
\label{fig:line3}
\end{figure*}

\section{Line contribution to the bolometric data at 870\,$\mu$m}\label{app:linecontamination}

\begin{figure*}[!t]
  \centering
      \begin{tabular}{lccc}
\rotatebox{90}{\parbox{2cm}{\centering $\Delta$ Dec.($\arcsec$) }}
&\includegraphics[clip, trim=0.0cm 0.0cm 0.0cm 0.0cm,height=6cm]{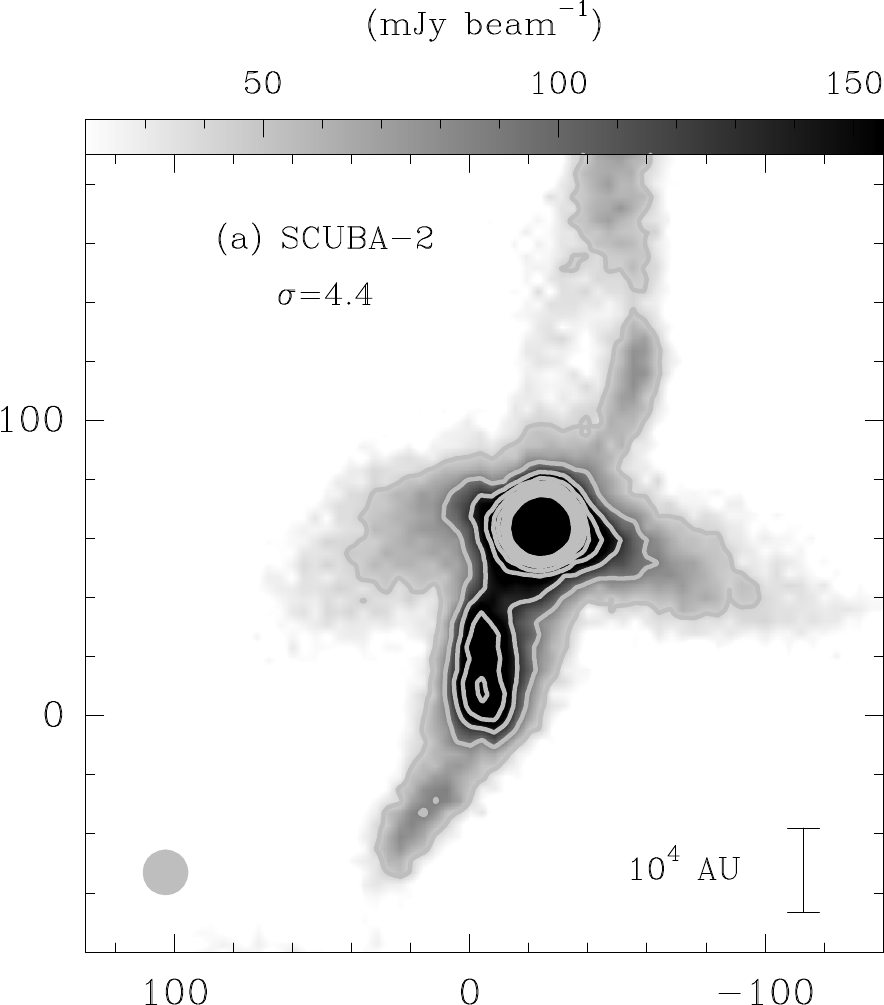}
&\includegraphics[clip, trim=1.1cm 0.0cm 0.0cm 0.0cm,height=6cm]{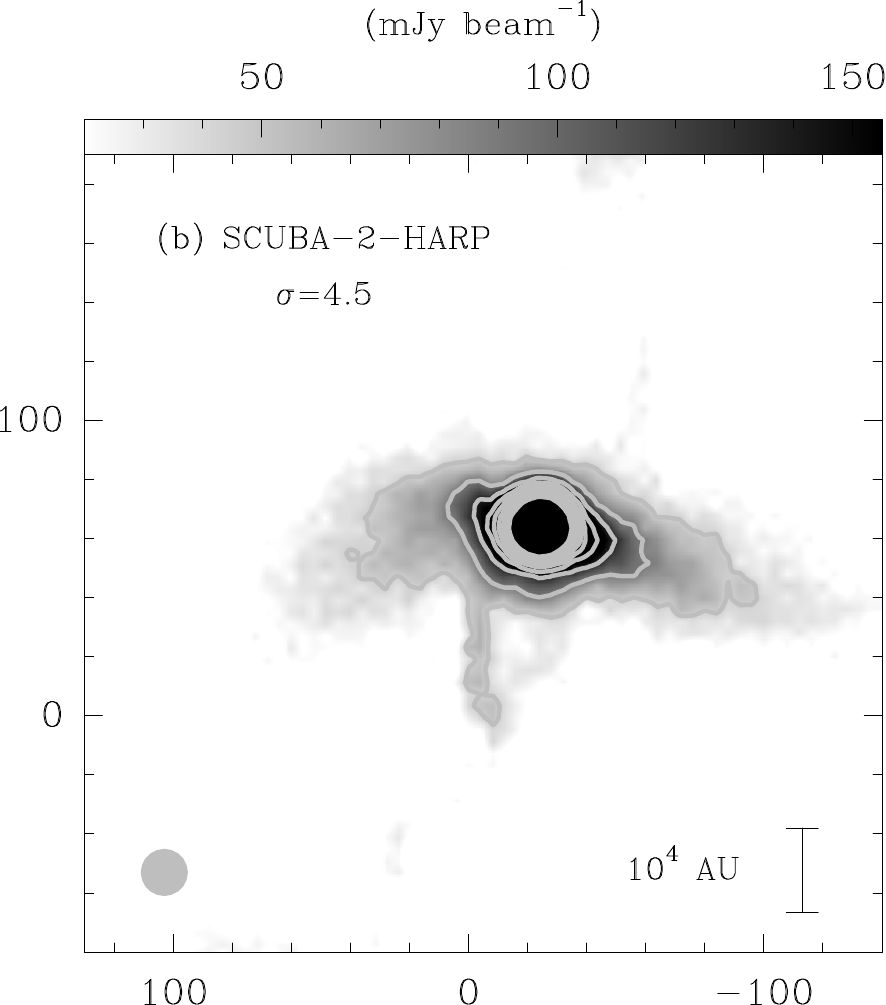}
&\includegraphics[clip, trim=1.1cm 0.0cm 0.0cm 0.0cm,height=6cm]{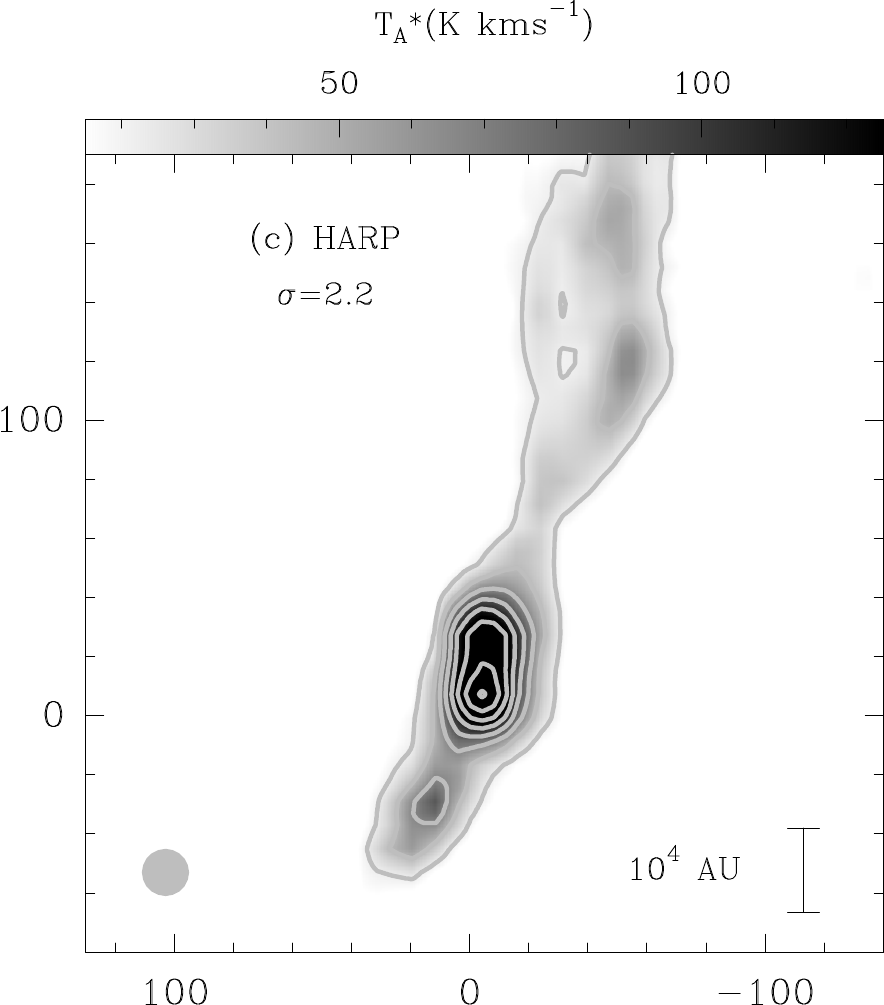}\\
&\includegraphics[clip, trim=0.0cm 0.0cm 0.0cm 0.0cm,height=6cm]{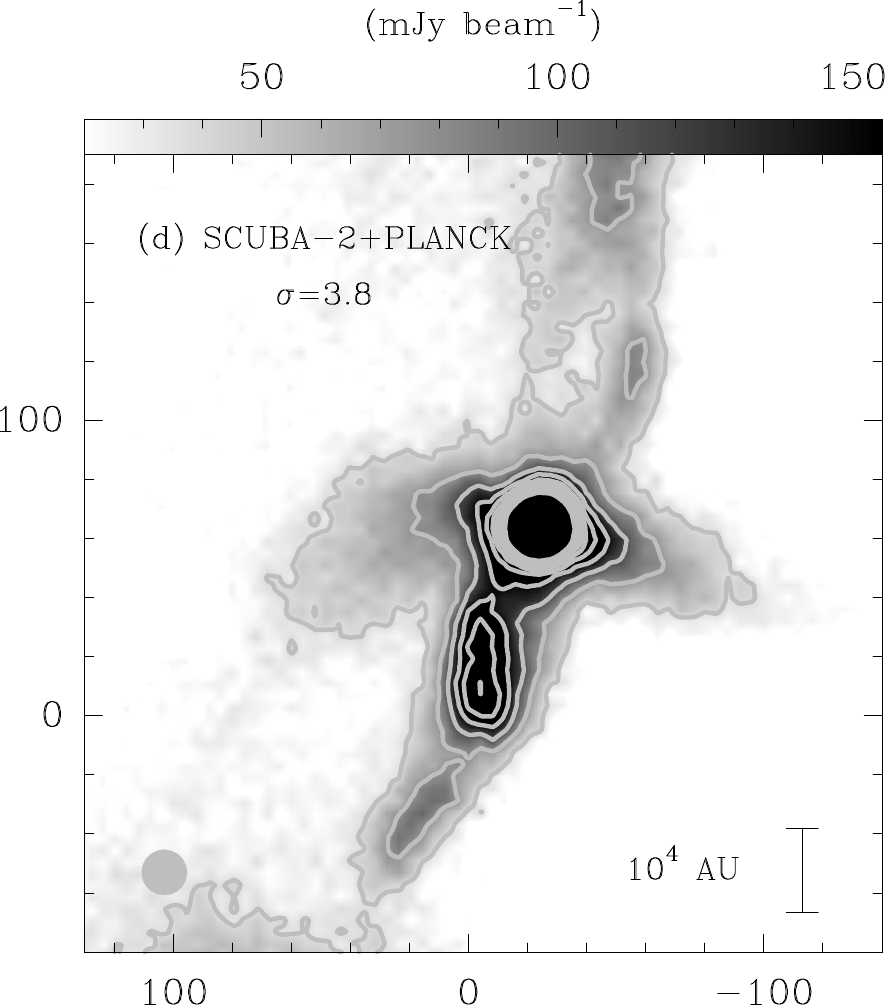}
&\includegraphics[clip, trim=1.1cm 0.0cm 0.0cm 0.0cm,height=6cm]{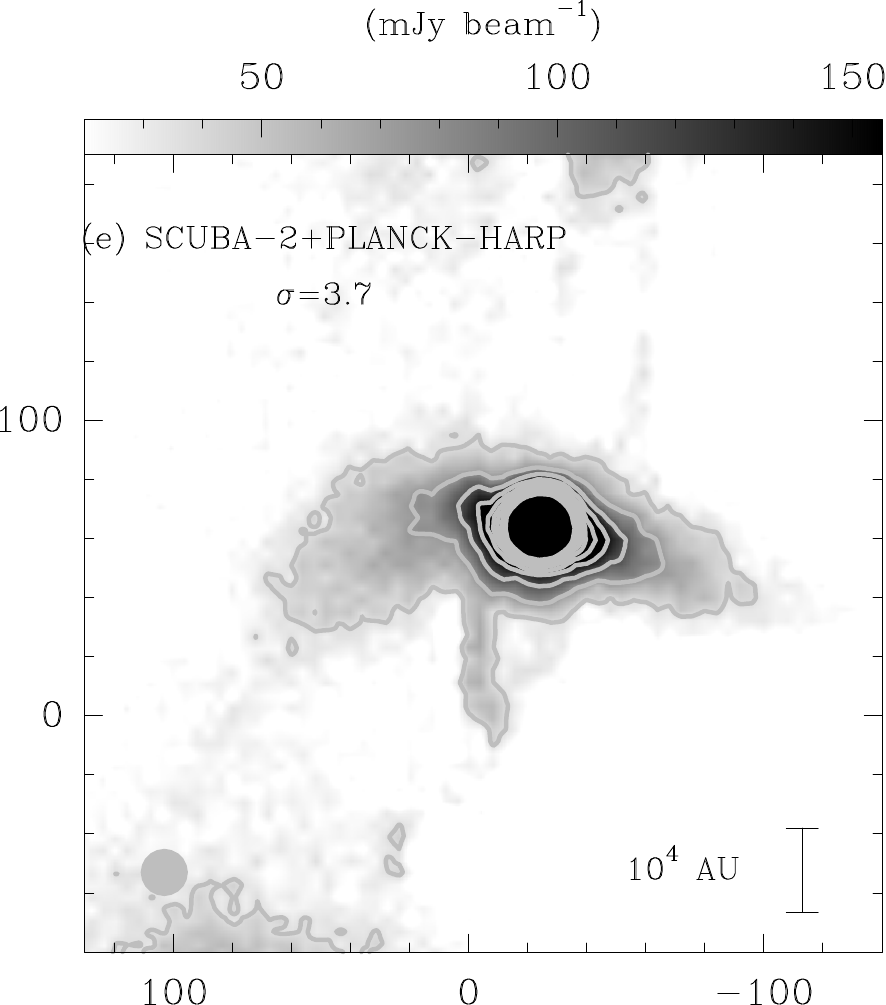}
&\includegraphics[clip, trim=1.1cm 0.0cm 0.0cm 0.0cm,height=6cm]{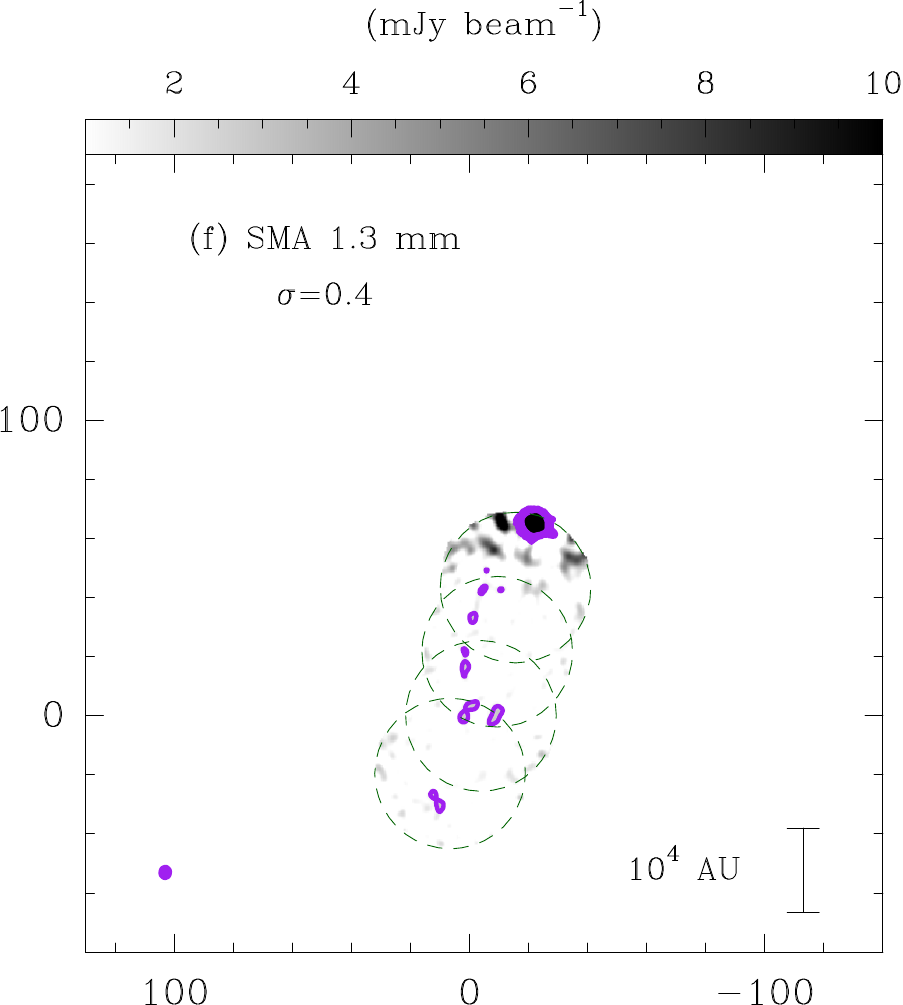}\\
& &$\Delta$R.A.($\arcsec$) &
\end{tabular}

\caption{Bolometer and dust continuum maps at 0.85\,mm and 1.3\,mm.
(a) Bolometric map obtained with JCMT/SCUBA-2 at 0.85\,mm.
(b) JCMT/SCUBA-2 0.85\,mm map after subtraction of molecular-line contamination estimated from JCMT/HARP observations.
(c) Integrated intensity map, in antenna temperature ($T_A^*$), of all lines observed with JCMT/HARP  at 0.85\,mm.
(d) Bolometric map obtained by combining JCMT/SCUBA-2 with \textit{Planck} data to recover spatially filtered emission.
(e) The combination map of JCMT/SCUBA-2 and \textit{Planck} after subtraction of molecular-line contamination.
(f) Dust continuum map derived from line-free channels of the SMA four-point mosaic observations at 1.3\,mm.
All panels are shown with the same reference position and coordinate range. The (0,0) offset corresponds to the phase center adopted in the NOEMA and VLA single-pointing observations of \citet{feng20a,feng22}: $\rm RA = 20^h39^m10^s.200$, $\rm Dec = +68^{\circ}01\arcmin10.50\arcsec$ (J2000).
Gray contours in panels (a)--(e) show emission levels from $10\sigma$ to $100\sigma$ in steps of $10\sigma$, where the $1\sigma$ rms values are given in each panel and have the same units as the color scale. 
Purple contours in panel (f) denote the SMA 1.3\,mm dust continuum, starting at $5\sigma$ and increasing in steps of $5\sigma$; the corresponding $1\sigma$ rms is also indicated.
In the lower left corner of each panel, the JCMT beam is shown as a gray filled circle, while the synthesized SMA beam is shown as a purple filled ellipse. 
In panel (f), green dashed circles indicate the four SMA mosaic pointings, with diameters equal to the SMA primary-beam FWHM at 1.3\,mm. Artifacts near the central protobinary system mm  in the northwestern mosaic field, caused by noise and/or missing short-spacing information, are masked in the contours but retained in the grayscale image.
}
\label{fig:jcmt-sma-combination} 
\end{figure*}

In Figure~\ref{fig:jcmt-sma-combination}, we present the 0.85\,mm bolometric maps obtained with JCMT/SCUBA-2 at an angular resolution of 14.5\arcsec, both before and after combining with \textit{Planck} observations to recover spatially filtered emission, as well as before and after removing the molecular-line contribution estimated from JCMT/HARP observations. We also show the 1.3\,mm dust continuum map derived from line-free SMA channels at an angular resolution of $\sim$3.4\arcsec, which exhibits a morphology consistent with that of the line-subtracted 0.85\,mm dust emission.

Table~\ref{tab:contamination} summarizes the contributions of molecular-line emission to the 0.85\,mm bolometric flux at the positions labeled in Figure~\ref{fig:source}.

\begin{table*}[!t]
\small
\caption{Line contamination factor estimated from Figure~\ref{fig:source} at 0.85\,mm.
}\label{tab:contamination}
\centering
\begin{tabular}{l|c|cc|cc|cc}
\hline
\hline
 Positions$^a$ &Bolometer$^b$ &\multicolumn{2}{c|}{CO\,(3\text{--}2)} &\multicolumn{2}{c|}{Other lines} &$100\times\frac{F_{\rm line}}{F_{\rm bolo}}$ &$100\times\frac{F_{\rm line}}{F_{\rm bolo}-F_{\rm line}}$ \\
 &(mJy\,beam$^{-1}$) &(K km\,s$^{-1}$)$^c$ &(mJy\,beam$^{-1}$)$^d$ &(K km\,s$^{-1}$)$^c$  &(mJy\,beam$^{-1}$)$^d$ &(\%)$^e$ &(\%)$^f$ \\
\hline

mm   &1232.6	 & 45.2   & 31.2      & 3.2      & 2.2      &  2.7    & 2.8 \\
B0n1 &129.2      & 68.3   & 47.1      & 22.4     & 15.6     &  48.5   & 94.1 \\
B0a  &154.0      & 92.3   & 63.7      & 57.1     & 39.4     &  66.9   & 61.8 \\
B0c  &214.5      & 139.1  & 96.0      & 79.0     & 54.5     &  70.1   & 234.7 \\
B0e  &198.6      & 139.0  & 95.9      & 73.2     & 50.5     &  73.7   & 280.6 \\
B1a  &199.7      & 143.2  & 98.8      & 80.0     & 55.2     &  77.1   & 337.1 \\
B1b  &187.1      & 123.4  & 85.1      & 68.1     & 47.0     &  70.6   & 240.2 \\
B1c  &125.7      & 106.0  & 73.2      & 44.7     & 30.9     &  82.8   & 481.1 \\
B1f  &215.2      & 154.6  & 106.7     & 90.5     & 62.4     &  78.6   & 367.0 \\
B1i  &99.3       & 81.4   & 56.2      & 29.7     & 20.5     &  77.2   & 339.0 \\
B2a  &92.0       & 64.6   & 44.6      & 45.5     & 31.4     &  82.6   & 473.4 \\
B2b  &93.4       & 46.0   & 31.8      & 26.0     & 17.9     &  53.1   & 113.3 \\

\hline
\hline
\multicolumn{8}{l}{Note. (a) At an angular resolution of 14.5\arcsec, B1a, B1b, and B1f are blended, B1c and B1i are blended.}\\
\multicolumn{8}{l}{\hspace{2.4em}(b) The bolometric flux is measured from the {0.85\,mm map constructed by combining the JCMT/SCUBA-2 and \textit{Planck} data.}}\\
\multicolumn{8}{l}{\hspace{2.4em}(c) The line integrated intensities in brightness temperature ($T_{\rm B}$) are converted from the antenna temperature ($T_A^*$) using $T_{\rm B}=T_A^*/0.71$.}\\
\multicolumn{8}{l}{\hspace{2.4em}(d) The line integrated intensities in flux density are derived from $T_{\rm B}$ under the Rayleigh-Jeans approximation.}\\
\multicolumn{8}{l}{\hspace{2.4em}(e) The line contribution to the bolometric flux.}\\
\multicolumn{8}{l}{\hspace{2.4em}(f) The percentage line contamination, defined as the integrated line emission divided by the line-subtracted bolometric flux.} \\
\end{tabular}
\end{table*}

\section{Two-component SED fitting}\label{app:fit}

Using  line-subtracted \textit{Herschel} data at 70, 160, 250, and 350\,$\mu$m, together with  combined JCMT and \textit{Planck} data (also line-subtracted) at 850\,$\mu$m, 
we performed SED fitting to derive the dust temperature, $\rm H_2$ column density, and dust opacity index $\beta$ across the entire field. We adopted a gas-to-dust mass ratio of 100 and assumed that the warm and cold dust components share the same $\beta$. Following the method of \citet{lin16,lin17,jiao22}, the two-component graybody fitting was carried out through the following steps:

(1) Masking: Data points with observed intensities below $\rm 3\sigma$ were excluded;

(2) Initial fit: A single-temperature SED fit was fitted to the 250, 350, and 850\,$\mu$m data using nonlinear curve fitting;

(3)  Extrapolation and constraint: The resulting parameters were extrapolated to 70 and 160\,$\mu$m,  where the observed fluxes or 3$\sigma$ upper limits provided additional constraints;

(4) Two-temperature fit: When the 70\,$\mu$m emission exceeded  the single-temperature model prediction by more than $3\sigma$, a two-temperature model was adopted to derive the warm dust component.

\begin{figure}[!ht]
  \centering
        \begin{tabular}{lc}
        \rotatebox{90}{\parbox{8cm}{\centering $\Delta$ Dec.($\arcsec$) }}
&\includegraphics[width=0.4\textwidth]{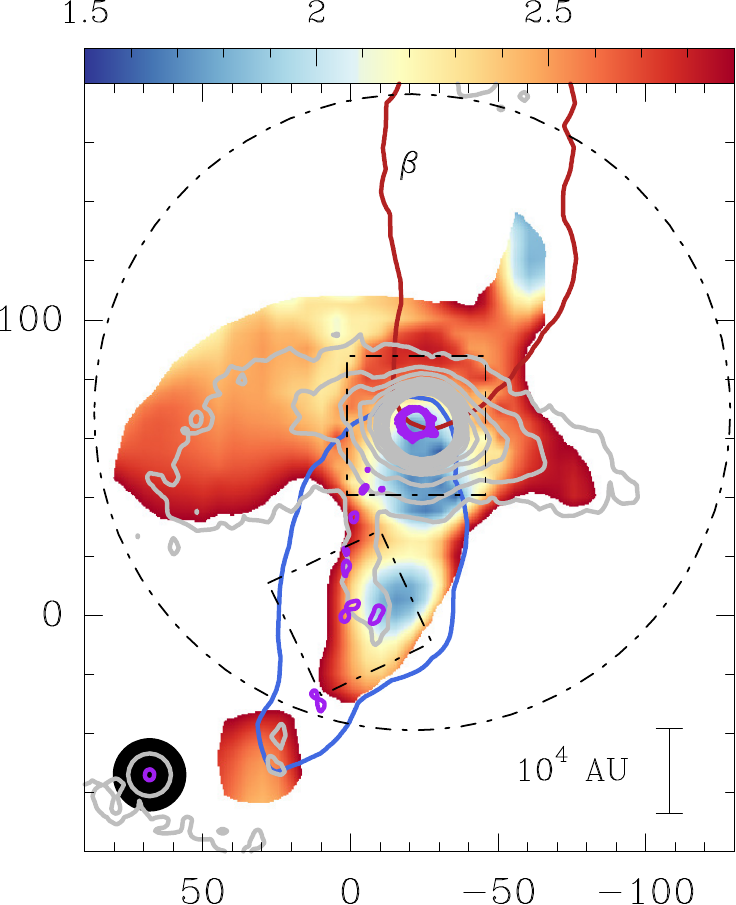}\\
&{~~~~~~~~~~~$\Delta$R.A.($\arcsec$) }\\
\end{tabular}
\caption{Map of the dust opacity index, $\beta$, derived from SED fitting to the line-subtracted bolometric maps from \textit{Herschel} (70--350\,$\mu$m) and the JCMT-\textit{Planck} combined data (850\,$\mu$m).  
The fitting adopts  a two-component model and assumes a gas-to-dust mass ratio of 100. 
Gray contours show the line-contamination-corrected 850\,$\mu$m continuum emission from the JCMT-\textit{Planck} combination.
while the red- and blue-shifted outflow lobes traced by JCMT CO\,(3\text{--}2) emission are shown in red and blue contours, respectively.
Purple contours represent the SMA 1.3\,mm dust continuum derived from line-free spectral channels. Contour levels are identical to those in Figure~\ref{fig:source}.
The gray circle in the lower left corner of each panel denotes the JCMT beam (14.5\arcsec), the purple ellipse marks the SMA synthesized beam ($\rm \sim3.4\arcsec$), and the black filled circle indicates the common angular resolution (24.9\arcsec) used for the parameter maps. 
The black dashed circle indicates the region where spectroscopic products from the COPS survey \citep{yang18} are used to remove line contamination from the 250\text{--}350\,$\mu$m bolometric maps. 
The black dashed boxes mark the protobinary envelope region covered by the CDF survey \citep{green16} and the B1 shocked region covered by the CHESS survey \citep{codella10,lefloch10}, where spectroscopic products are used to subtract line emission from the 70 and 160\,$\mu$m bolometric maps. 
Regions where the SED fitting uncertainties exceed 50\% of the derived values, where the dust opacity index $\beta$ could not be constrained, and where  the smoothed 850\,$\mu$m emission falls below the $\rm 3\,\sigma$ level, are masked.
} 
\label{fig:beta}
\end{figure}

Figure~\ref{fig:sed-point} presents representative examples of the SED fits ({solid lines}), including the derived temperatures of the warm and cold components, the total $\rm H_2$ column density, and the corresponding dust opacity index $\beta$ at four selected positions (Table~\ref{tab:clump}). 
{For comparison, the fits obtained using the original bolometric data before molecular-line contamination correction are shown as dashed lines, illustrating the impact of line contamination on the derived fitting parameters.}
 The resulting spatial distribution of $\beta$ is presented in Figure~\ref{fig:beta}.

\section{The gas-grain forming path of $\rm NH_3$ constrained by observations}\label{app:model}

\subsection{Physical model}\label{app:phy}
The physical model adopted for each region consists of two evolutionary stages: a pre-shock stage and a shock stage. The pre-shock stage represents relatively quiescent molecular gas with physical conditions characteristic of the protostellar envelope. During this phase, the gas temperature is fixed at 10\,K, and the total hydrogen nuclei density is set to $n_\mathrm{H} = 1\times10^5$ cm$^{-3}$.

{During} the shock stage, gas parcels are assumed to be ejected sequentially from the protobinary system and propagate downstream along the outflow. Following the observational analysis of \citet{podio16}, and adopting the same relative chronology after scaling the dynamical timescales to the updated distance, the projected separations between {successive} parcels are converted into their corresponding ejection time intervals.
Taking the ejection of  {the first} parcel from the protobinary system as $t=0$, it reaches the projected position observed today as B2a after $\sim2500$\,yr, {where it drives} a shock into the local gas. For comparison, we consider two additional parcels ejected at $t=750$\,yr and $t=1600$\,yr. 
After travel times of 1750\,yr and 900\,yr, respectively, these parcels reach the projected positions {observed today} as B1c and B0n1, {where they generate shocks at the current observational epoch.}

\begin{figure*}[!t]
\includegraphics[width=1.0\textwidth]{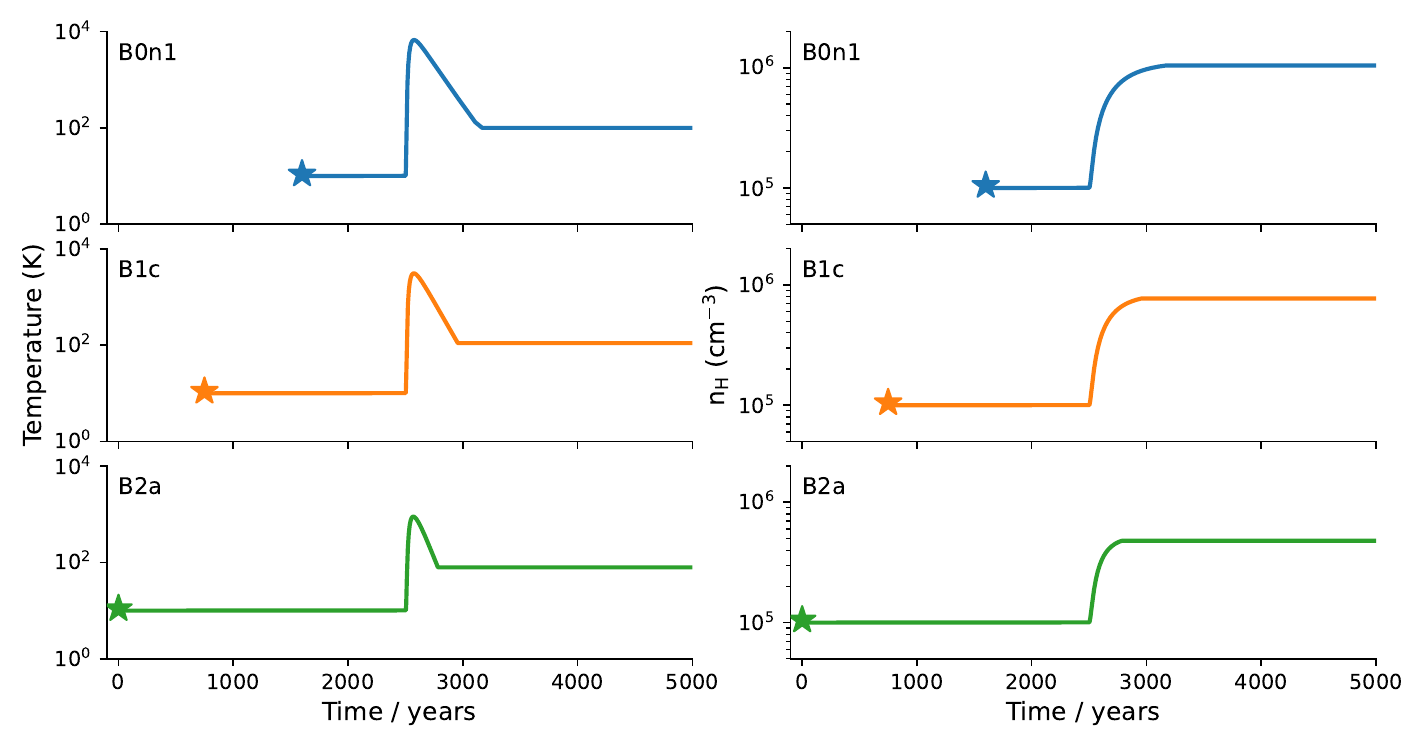}
\caption{
Gas temperature and hydrogen nuclei density as functions of time at positions B0n1, B1c, and B2a, as adopted in the chemical models. Star symbols indicate the ejection times of individual parcels from the protostar; physical parameters are not shown prior to each parcel ejection.
}
\label{fig:physicalchanges}
\end{figure*}

Figure~\ref{fig:physicalchanges} shows the temporal evolution of the gas temperature and density at each projected position {following} the arrival of the corresponding gas parcel. To {highlight} the time intervals between successive ejection events, we do not show the quiescent phase prior to ejection, during which the gas remains unaffected by outflow-driven shocks.
Within this framework, the gas parcels currently observed at different projected positions {are assumed to have} reached their respective locations at approximately the same observational epoch, with timing uncertainties that are small compared with their overall travel times. Consequently, we do not attempt to constrain the delay between the onset of shock-driven chemistry and the time of observation. Instead, following \citet{lefloch21}, we adopt a monotonic decrease in shock velocity along the flow,  from 60\,$\rm km\,s^{-1}$ at B0n1 to 40\,$\rm km\,s^{-1}$ in B1c and 20\,$\rm km\,s^{-1}$ at B2a.

Various shock models  (J-type, C-type, $\rm C^*$-type, and CJ-type) have been explored in the literature for the B1 and B2 regions \citep[e.g.,][]{burkhardt19,james20,baijot25}. However, these studies generally find no strong chemical differentiation among shock types for the species considered \citep{james20}. We therefore follow \citet{viti11} and adopt C-type shock models {established in \textsc{UCLCHEM} \citep{holdship17}} for all three regions.

The resulting post-shock hydrogen nuclei densities are $1\times10^{6}$\,cm$^{-3}$, $7\times10^{5}$\,cm$^{-3}$, and $4\times10^{5}$\,cm$^{-3}$ for B0n1, B1c, and B2a, respectively.  
The post-shock gas temperatures are adopted from observational constraints, with $\gtrsim$100\,K in B0n1, $\lesssim$80\,K in B2a, and an intermediate value of $\sim110$\,K in B1c. 
{While varying between locations to reflect the global shock gradients, the temperature is assumed to remain stationary for any given shock layer during its subsequent chemical evolution.}
 
\subsection{Chemical model}\label{app:che}

As shown in Figure~\ref{fig:source}, the dust continuum emission toward B1c is systematically weaker than that toward B0n1, both on the  850\,$\mu$m image {at a linear resolution of 4930\,au} and on the 1.3\,mm image {at a linear resolution of 1200\,au}. In contrast, the $\rm NH_3$ line emissions toward B1c is significantly stronger, corresponding to a higher $\rm NH_3$ column density than that toward B0n1 (Figure~\ref{fig:nh2-sma}; {see also} \citealt{feng22}). Owing to missing flux, B2a may not represent the true peak of the dust continuum emission. Nevertheless, we included B2a in our analysis as an intermediate case, with an $\rm NH_3$ abundance between those of B1c and B0n1.

Based on the two-stage physical model described in Appendix~\ref{app:phy}, we performed the chemical modeling using the \textsc{Nautilus} code \citep{ruaud16,wakelam24}. The chemical reaction network and physical processes follow the default configuration of the package. The chemical abundances obtained at the end of the pre-shock stage ($\sim10^{6}$\,yr; Table~\ref{tab:abundances}) are adopted as the initial conditions for the subsequent outflow–shock stage.

\begin{figure*}[!t]
\includegraphics[width=1.0\textwidth]{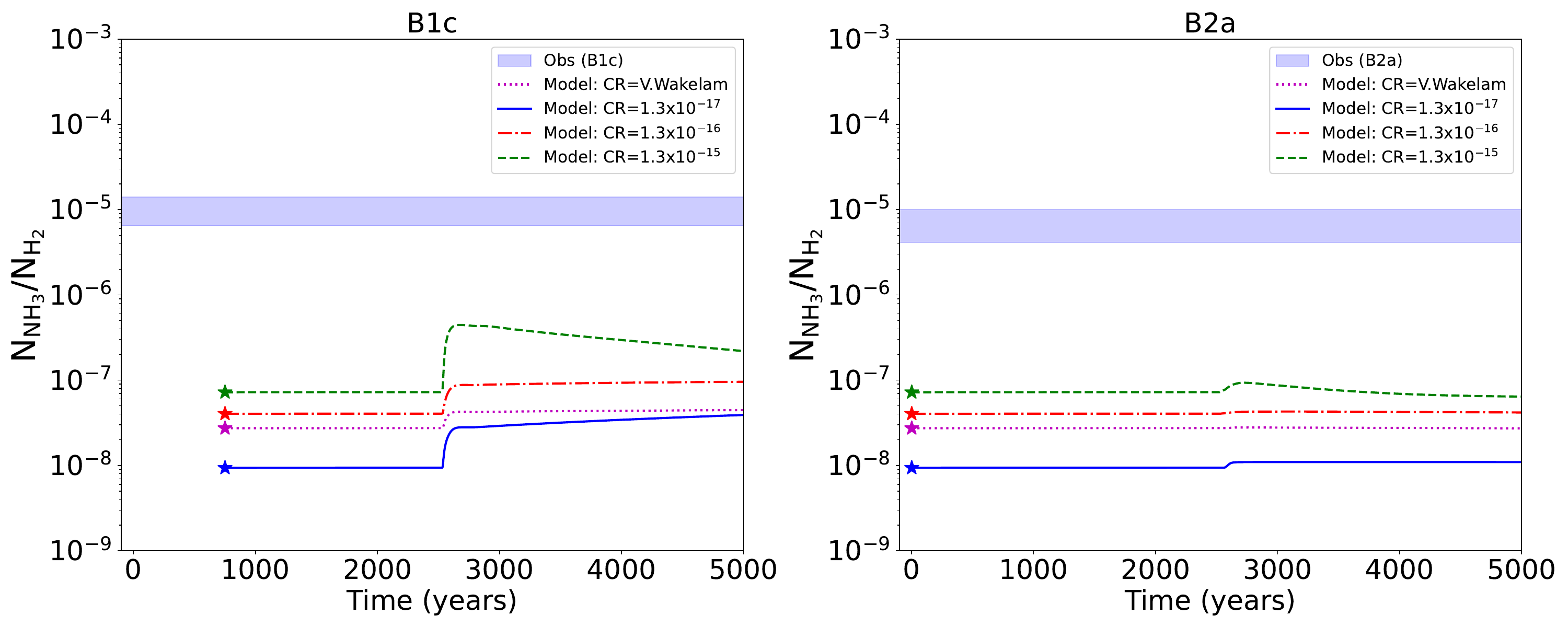}
\caption{
Time evolution of the gaseous $\rm NH_3$ abundance relative to $\rm H_2$ toward the B1c (left) and B2a (right) regions, as predicted by pure gas-phase chemical models.
Each curve corresponds to a different assumed {CRIR} (in unit of $\rm s^{-1}$), adopting a gas-to-dust mass ratio of 100.
The blue shaded regions indicate the observed abundance ranges; the upper and lower bounds reflect uncertainties associated with assuming gas-dust thermal coupling (i.e., equal gas and dust temperatures) and a dust temperature of 50\,K.
Star symbols indicate the ejection times of individual parcels from the protostar; gaseous $\rm NH_3$ abundance are not shown prior to each parcel ejection.
}
\label{fig:gasNH3_diff_CR}
\end{figure*}

We explored two classes of chemical models {to evaluate the relative contributions of gas-phase and grain-surface pathways. As a benchmark for our analysis, we first establish a pure gas-phase model to quantify the extent to which gas-phase reactions can account for the observed abundances. In this case, $\rm NH_3$ is formed primarily through the dissociative recombination of $\rm NH_4^+$ ($\rm NH_4^+ \rightarrow NH_3$).
In this initial model, grain-surface and ice chemistry are intentionally neglected to isolate the gas-phase efficiency, with the {CRIR}  treated as the primary adjustable parameter.} 
We test constant ionization rates of $1.3\times10^{-17}$, $1.3\times10^{-16}$, and $1.3\times10^{-15}$\,s$^{-1}$, as well as a time-dependent rate computed following \citet{wakelam21}, applied to both the pre-shock and shock stages. Although higher ionization rates systematically enhance the gas-phase $\rm NH_3$ abundance, even the extreme value of $1.3\times10^{-15}$\,s$^{-1}$ yields peak abundances that remain one to two orders of magnitude below the observed values (Figure~\ref{fig:gasNH3_diff_CR}). 
These results provide empirical verification that gas-phase chemistry alone is insufficient, reinforcing the necessity of incorporating grain-surface processing and subsequent desorption to reproduce the $\rm NH_3$ enhancement in L1157-B1.

\begin{table}[!t]
\small
\caption{Initial abundances of elements with respect to total hydrogen nuclei. }
\label{tab:abundances}
\centering
 \scalebox{1}{
\begin{tabular}{cp{2.5cm}}
\hline
\hline
      Species &Abundance\\

\hline    
 He       & $9.0 \times 10^{-2}$ \\
 N        & $6.2 \times 10^{-5}$\\
 O        & $2.4 \times 10^{-4}$\\
 H$_2$    & $5.0 \times 10^{-1}$ \\
 C$^{+}$  & $1.7 \times 10^{-4}$\\
  S$^{+}$  & $8.0 \times 10^{-8}$\\
  Si$^{+}$ & $8.0 \times 10^{-9}$\\
  F & $6.68 \times 10^{-9}$\\
  Na$^{+}$ & $2.0 \times 10^{-9}$\\
  Mg$^{+}$ & $7.0 \times 10^{-9}$\\
  Fe$^{+}$ & $3.0 \times 10^{-9}$\\
  P$^{+}$  & $2.0 \times 10^{-10}$\\
  Cl$^{+}$ & $1.0 \times 10^{-9}$\\                                                      
\hline
\hline

\end{tabular}
}
\end{table}

Next, we adopted a three-phase chemical model that includes gas-phase, grain-surface, and ice-mantle chemistry. In this framework, the chemical evolution of all three regions is followed starting from $t = 0$, defined as the ejection time of the gas parcel currently observed at B2a. The {CRIR} in both evolutionary stages is treated as time-dependent and computed following \citet{wakelam21}, while dust-related processes play a central role.

During shock propagation, we explicitly included shock-induced sputtering as an efficient nonthermal desorption mechanism capable of releasing grain-surface species into the gas phase, even at relatively low dust temperatures \citep{caselli97}. 
{In the present work, we adopt a simplified zero-dimensional treatment in which grain-mantle material is injected into the gas phase instantaneously at the onset of the shock. This approximation is motivated by previous shock models \citep{jimenez08,lefloch21}, which show that sputtering timescales are typically on the order of a few years and generally shorter than $\sim10$\,yr, substantially smaller than the chemical and dynamical timescales considered in our models ($\gtrsim10^{3}$\,yr). }
 The gas and dust temperatures are assumed to be decoupled (i.e., in all three regions), the dust temperature is assumed to increase from 10\,K to 50\,K due to shock heating, while the gas temperature is computed self-consistently within the shock model \citep{holdship17}.

\end{appendix}

\end{document}